# Descriptive and Foundational Aspects of Quantum Cognition

Reinhard Blutner[a] & Peter beim Graben[b]

[a]ILLC, Universiteit van Amsterdam, Amsterdam, The Netherlands, 1090 GE.
[b]Department of German Studies and Linguistics, Humboldt-Universität zu Berlin, 10099 Berlin, Germany.

**Abstract:**
Quantum mechanics emerged as the result of a successful resolution of stringent empirical and profound conceptual conflicts within the development of atomic physics at the beginning of the last century. At first glance, it seems to be bizarre and even ridiculous to apply ideas of quantum physics in order to improve current psychological and linguistic or semantic ideas. However, a closer look shows that there are some parallels in the development of quantum physics and advanced theories of cognitive science dealing with concepts, conceptual composition, vagueness, and prototypicality. In both cases of the historical development, the underlying basic ideas are of geometric nature. In psychology, geometric models of meaning have a long tradition. However, they suffer from many shortcomings which are illustrated by discussing several puzzles of bounded rationality. The main suggestion of the present approach is that geometric models of meaning can be improved by borrowing basic concepts from (von Neumann) quantum theory.

In the first part of this article, we consider several puzzles of bounded rationality. These include the Allais- and Ellsberg paradox, the disjunction effect, the conjunction and disjunction fallacies, and question order effects. We argue that the present account of quantum cognition – taking quantum probabilities rather than classical probabilities – can give a more systematic description of these puzzles than the alternate and rather eclectic treatments in the traditional framework of bounded rationality. Unfortunately, the quantum probabilistic treatment does not always and does not automatically provide a deeper understanding and a true explanation of these puzzles. One reason is that quantum approaches introduce additional parameters which possibly can be *fitted* to empirical data but which do not necessarily *explain* them. Hence, the phenomenological research has to be augmented by responding to deeper foundational issues.

In the second part of this article, we aim to illustrate how recent progress in the foundation of quantum theory can help to answer the foundational questions of quantum cognition. This includes the opportunity of interpreting the free parameters, which are pure stipulations in the quantum probabilistic framework. Making a careful distinction between foundational and phenomenological research programs, we explain the foundational issue from the perspective of Piron, Foulis, Randall, and others, and we apply it to the foundation of quantum cognition. In this connection, we show that quantum probabilities are of (virtual) conceptual necessity if grounded in an abstract algebraic framework of orthomodular lattices. This framework is motivated by assuming partial Boolean algebras (describing particular perspectives) that are combined into a uniform system while considering certain capacity restrictions. It is at this point that one important aspect of the whole idea of bounded rationality directly enters the theoretical scenery of quantum cognition: resource limitation. Another important aspect of the foundational issue is that it automatically leads to a distinction between probabilities that are defined by pure states and probabilities arising from the statistical mixture of such states. It is possible to relate this formal distinction to the deep conceptual distinction between risk and ignorance. A third outcome is the possibility to identify quantum aspects in dynamic macro-systems using the framework of symbolic dynamics, closely related to the operational perspective of Piron, Foulis, Randall, and others. This helps to understand the ideas of epistemic complementarity and entanglement, and to analyse quantum aspects in third generation neural networks.

# 1. Introduction

In a series of recent papers, quantum probabilities are discussed as providing an alternative to classical probability for understanding cognition (e.g., Aerts 2009; Aerts et al. 2005; Blutner 2009; Blutner et al. 2013; Bruza et al. 2009; Busemeyer and Bruza 2012; Busemeyer et al. 2006; Conte et al. 2008; beim Graben 2004; Gabora and Aerts 2002; Kitto 2008; beim Graben 2014). In considerable detail, they point out several cognitive phenomena of perception, decision and reasoning that cannot be explained on the basis of classical probability theory, and they demonstrate how quantum probabilities can account for these phenomena. An obvious way to downplay this chain of arguments is by demonstrating that besides classical probability and quantum probability models, alternative approaches are possible that could also describe the phenomena under study without using the strange and demanding instruments of quantum probability. This often happens in the scientific context of bounded rationality where alternative proposals are developed. For instance, one could argue that the so-called conjunction puzzle can be resolved by simple heuristics (Gigerenzer 1997), and the well-known question ordering effects by query theory (Johnson et al. 2007).

A general strategy to encounter such criticism is to look for a universal motivation of quantum probabilities which is based on fundamental (architectural) properties of the area under investigation. As Kuhn (1996) clarified, such basic assumptions constituting a theoretical paradigm normally cannot be justified empirically. Basic assumptions which concern the general architecture of the theoretical system are referred to as *design features*. Using a term that is common in the generative linguistic literature (Chomsky 1995, 2005), we shall call properties that are consequences of such design features as applying with (*virtual*) *conceptual necessity*. Hence, a central aim of this article is to demonstrate the conceptual necessity of the basic formalism of quantum mechanics for modelling concepts and their combination (including a description of vagueness and typicality).

In the present literature, there are several approaches that seek for a general justification of quantum probabilities in the context of cognitive science. For example, Kitto (2008) considers very complex systems such as the growth and evolution of natural languages and other cultural systems and argues that the description of such systems cannot be separated from their context of interaction. She argues that quantum interaction formalisms provide a natural model of these systems "because a mechanism for dealing with such contextual dependency is inbuilt into the quantum formalism itself" (Kitto 2008: p. 12). Hence, the question of why quantum interaction is necessary in modelling cognitive phenomena is answered by referring to its nature as a complex epistemic system. A related approach is taken by Aerts and Sassoli de Bianchi (2014a, 2014b). These authors model the situation typical for measurements in cognitive science and argue that the observer's lack of information relates to uniform and non-uniform fluctuations generated on the measurement situation. The emergence of quantum probabilities can be handled then by a first-order approximation of a more general non-uniform quantum approach.

In their recent book, Busemeyer and Bruza (2012), give several arguments why quantum models are necessary for cognition. Some arguments relate to the cognitive mechanism of judgments. Judgments normally do not take place in definite situations. Rather, judgments create the context where they take place. This is the dynamic aspect of judgments also found in dynamic models of meaning (beim Graben 2014). Another is the logical aspect. The logic of judgments does not obey classical logic. Rather, the underlying logic is very strange with asymmetric conjunction and disjunction operations. When it comes to considering probabilities and conditioned probabilities the principle of *unicity* is violated, i.e. it is impossible to assume a single sample space with a fixed probability distribution for judging all possible events. Other arguments relate to the problem of compositionality in cognitive semantics (Blutner et al. 2003; Blutner et al. 2004; de Hoop et al. 2007; Spenader and Blutner 2007). In Sections 2 - 4 we will discuss some of these arguments.

Another line of argumentation seeks to answer the question of "why quantum models of cognition" by speculating about implications for brain neurophysiology. In the taken algebraic approach, even classical dynamical systems such as neural networks, could exhibit quantum-like properties in the case of coarse-graining measurements, when testing a property cannot distinguish between epistemically equivalent states (beim Graben 2004). Beim Graben & Atmanspacher (2009) used this "epistemic quantization" for proving the possibility of incompatible properties in classical dynamical systems. In neuroscience, most measurements, such as electroencephalography or magnetic resonance imaging, are coarse-grainings in this sense. Thus, the quantum approach to cognition has direct implications for brain neurophysiology, without needing to refer to a "quantum brain", as recently indicated by Pothos & Busemeyer (2013). A novel application of this idea using Hebbian neurodynamics as an underlying classical system to describe emerging properties that exhibit quantum-like traits is given by Acacio de Barros and Suppes (2009), Large (2010), and by others.

As far as we can see, none of the foundational programs of grounding quantum cognition explicitly refers to the logical programs of theoretical physics that aim to ground all physical processes using the perspective of "operational realism". In the following, we will come with a new proposal of answering the why question of quantum cognition, one that is based on operational realism. Mainly the program developed by the "Geneva school" (Jauch 1968; Piron 1972, 1976) is of special interest here:

> According to these authors, a "rational" reconstruction of Quantum Mechanics should be stated in terms of the actual operational meaning of the fundamental quantum mechanical concepts and principles, i.e. in terms of outcomes of possible experiments ("answers" to "questions") that could in principle be performed on a given physical system. This approach might be called "operational-realist", to distinguish it from the "pure" operationalism of neo-positivists and instrumentalists (Baltag and Smets 2011: p. 285)

The basic idea of the Geneva school (borrowed from Mackey 1963) is to find a general and abstract formulation of quantum mechanics without explicitly starting from the structures of a Hilbert space. Instead, the idea is to justify the use of Hilbert space via a representation theorem (for the historical details, see Smets 2011). Further research by Foulis, Randall and colleagues (Foulis et al. 1983; Foulis and Randall 1972; Randall and Foulis 1973; Foulis 1999) substantiates this point.

Summarizing, the present article makes a careful distinction between phenomenological research and foundational research. In the first part of this article (Sections 2-4) we consider several puzzles of bounded rationality, and we argue that the present account of quantum cognition – taking quantum probabilities rather than classical probabilities – does not automatically provide a deeper understanding and a true explanation of these puzzles. The reason is that the quantum idea introduces several new parameters which possibly can be *fitted* to empirical data but which do not necessarily *explain* them. Hence, the phenomenological research has to be augmented by responding to deeper foundational issues. In the second part (Sections 5-7) of this article, we aim to illustrate how present progress in the foundation of quantum theory can help to answer the foundational questions of quantum cognition. This includes the opportunity of interpreting the free parameters which are pure stipulations in the quantum probabilistic framework.

Here is a more detailed outline of this article. In Section 2, we will review the basic idea of bounded rationality in cognitive science and we will outline prospect theory as a basic approach that makes use of this idea. Further, we make clear why most of this work can best be characterized as "phenomenological research", i.e. research that is related to the precise description of different phenomena without claiming a general, explanatory value. Section 3 introduces a broader series of puzzles that are discussed in the context of bounded rationality and argues that prospect theory and its follow-ups are of limited value only when it comes to ask for explanatory solutions. Section 4 introduces quantum probability as a new way to handle the puzzles and discusses the important role of interference effects. Unfortunately, we will see that within this field one and the same problem is often handled in different ways. In addition, the explanatory value of particular treatments is not always convincing. For that reason, it is essential to ask the foundational question, as done in the second part. This part starts with Section 5, which explains the basic distinction between foundational and phenomenological research programs. Further, this section makes a historical note about the uncertainty principle and it illustrates that there are many interpretations of complementarity in the historical context. It also argues for an epistemic interpretation of complementarity in quantum cognition. In Section 6, we explain the foundational issue from the operational perspective of Piron, Foulis, and Randall, and give a closely related formulation in terms of symbolic dynamics. Further, we relate bounded rationality with resource limitations and consider a particular application of the ideas developed so far to research in speed-accuracy trade-offs. Finally, in Section 7 we discuss some empirical consequences of the present foundational perspective and we draw some general conclusions.

## 2. Bounded rationality and cognitive science

Early models of cognitive science are based on general ideas of rationality as developed in logic, probability theory and decision theory. For example, researchers of cognitive science have borrowed the idea of rational decisions from mathematical economics and game theory (e.g., Von Neumann and Morgenstern 1944; Savage 1954). The classical idea is to assume (i) that rational agents are expected utility maximizers (using the mathematical conception of expected utility based on a classical probability measure), and (ii) that they maximize their expected utility with respect to the underlying economic model. Attempts to overcome the theoretical shortcomings of the classical attempt have resulted in the development of the bounded rationality project, originally proposed by Herbert Simon (1955). Simon claimed that we have to replace the "global rationality of the economic man" with a kind of behaviour that is compatible with the boundedness of the human decision maker and the particular kinds of environments in which such organisms exist. Boundedly rational agents experience limits in formulating and solving complex problems and in processing (receiving, storing, retrieving, transmitting) information. There is a number of dimensions along which "classical" models of rationality can be made more realistic without giving up rigorous formalization. A good example is prospect theory (Kahneman and Tversky 1979):

> In the classical theory, the utility of an uncertain prospect is the sum of the utilities of the outcomes, each weighted by its probability. The empirical evidence reviewed above suggests two major modifications of this theory: 1) the carriers of value are gains and losses, not final assets; and 2) the value of each outcome is multiplied by a decision weight, not by an additive probability. The weighting scheme used in the original version of prospect theory and in other models is a monotonic transformation of outcome probabilities. (Tversky and Kahneman 1992)

Formally, the difference between the classical scheme and prospect theory is shown here:

(1) a. $U(f) = \Sigma_i\, p_i \cdot u(x_i)$
    b. $U(f) = \Sigma_i\, \gamma(p_i) \cdot v(x_i)$

In the classical scheme of utility theory (1a), prospects or gambles $f = (x_1, p_1; \ldots, x_n, p_n)$ are contracts that yield (monetary) outcome $x_i$ with probability $p_i$ ($\Sigma_i\, p_i = 1$). The utility $u(x_i)$ is a direct expression of the monetary outcome $x_i$. The overall utility of a prospect $f$, denoted by $U(f)$, is the expected utility of its outcomes $x_i$.

It is convenient, to illustrate a prototypic decision problem as pictured in Figure 1. Depending on the chosen utility function, the figure illustrates both the classical scheme and prospect theory.

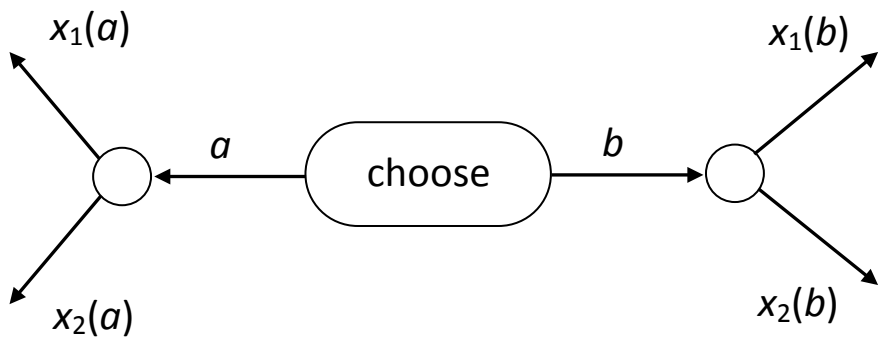

Figure 1: Prototypic decision problem. An agent has to decide between two prospects, *a* and *b*, leading to payoffs $x_1(a)$ and $x_2(a)$ for prospect *a* and $x_1(b)$ and $x_2(b)$ for prospect *b.* In classical decision theory, prospect *a* is chosen over prospect *b* if the classical utility U(*a*) is bigger than the classical utility U(*b*). Note that sometimes more than two payoffs are possible for a given prospect. And in some special cases, where only one possibility is open, one payoff is present only for one of the given prospects.

Prospect theory calculates the overall value of a prospect as shown in (1b). Hereby, the scale *v* assigns to each outcome $x_i$ a number which reflects the subjective value of that outcome utility[1]. The second scale, $\gamma$, associates with each probability $p_i$ a decision weight $\gamma(p_i)$, which reflects the impact of $p_i$ on the over-all value of the prospect. Than transformation $\gamma$ translates standard probabilities into numbers that do not represent a probability measure but still are normalized (i.e., $\gamma(0)=0$, $\gamma(1)=1$). The sum of $\gamma\,(p)$ and $\gamma\,(1-p)$ is typically less than unity. The function $\gamma$ is a nonlinear function which is formed by reference points eliciting the nonlinear deformations:

> For probability, there are two natural reference points—certainty and impossibility—that correspond to the endpoints of the scale. Therefore, diminishing sensitivity implies that increasing the probability of winning a prize by .1 has more impact when it changes the probability of winning from .9 to 1.0 or from 0 to .1 than when it changes the probability from,

---

[1] This is typically a exponential function of the outcome, and it takes reference points into account reflecting that losses hurt more than gains feel good – for instance by taking $v(x) = x^{.65}$ for $x \geq 0$ (gains) and $v(x) = -(-x)^{.75}$ for $x \leq 0$ (losses) (Tversky & Shafir, 1992, p. 308)

say, .3 to .4 or from .6 to .7. This gives rise to a weighting function that is concave near zero and convex near one. (Tversky & Fox, 1995, p. 748)

A simple example should illustrate the difference between standard utility theory and prospect theory (cf. Kahneman and Tversky 1979). Which of the following prospects would you prefer?

*a*: 50% chance to win 1,000, 50% chance to win nothing; i.e. *a* = (1000,.5; 0, .5).
In this case $U(a) = .5\, u(1000)$
*b*: 450 for sure; i.e. *b* = (450,1.). In this case $U(b) = u(450)$

In case of using classical utility theory, people should prefer prospect *a* over prospect *b* since in the first case the expected utility is higher. In other words, we expect $u(1000)/u(450) > 2$ assuming a linear function for u. Several forms of such questionnaires were constructed and in each case (with outcomes referring to Israeli currency) the majority of the people voted in the opposite way.

In order to analyse why classical decision theory fails we have to consider two other prospects with the same outcomes but with different probabilities:

*c*: 10% chance to win 1,000, 90% chance to win nothing; i.e. *c* = (1000,.1; 0, .9). In this case $U(c) = .1\, u(1000)$
*d*: 20% chance to win 450, 80 % chance to win nothing; i.e. *d* = (450, .2; 0, .8). In this case $U(d) = .2\, u(450)$

While in the former case, classical decision theory predicts that rational agents prefer *b* over *a* iff $u(1000)/u(450) < 2$, in the latter case the theory predicts that rational agents prefer *d* over *c* iff exactly the same condition holds (since the condition $0.2\, u(450) > .1\, u(1000)$ is equivalent to the former condition $u(1000)/u(450) < 2$). Interestingly, the majority of people who took place in corresponding experiments preferred *c* over *d* in the latter case. As a consequence, the classical utility function is not able to describe the empirical data even in case of using a nonlinear utility function *v*. Hence, the assumed probabilities have to be corrected, by taking the function $\gamma$ into account as stated in prospect theory. In order to account for the empirical data we have to state the following empirical condition: $\gamma(1)/\gamma(.5) > \gamma(.2)/\gamma(.1)$. Taking the reference points 0 and 1 into account as argued above, it is plausible to assume that $\gamma(1) = 1$ and $\gamma(.5) \approx .5$; hence, $\gamma(1)/\gamma(.5) \approx 2$. Further, because of the concavity of the $\gamma$-function near zero (effect of overweighting small probabilities) it is plausible that $\gamma(.2)/\gamma(.1) < 2$ and the empirically required condition is satisfied.

At this point it should be obvious how the framework of bounded rationality is put into action: People are assumed to evaluate their losses and gains and to judge their probabilities by assuming certain heuristics. The model is descriptive but not explanatory. It tries to model real life choices by introducing several heuristically motivated parameters that can be fitted with the empirical data. Looking at the huge literature of recent approaches to bounded rationality (e.g. Gigerenzer and Selten 2001) suggests an adaptive toolbox of different approaches and models to handle the list of phenomena headed under the same title. From an explanatory point of view such an approach is not really encouraging, and a more systematic approach would be highly welcome. Unfortunately, there is no consensus on the definition of bounded rationality or what the critical questions of the project are (Rubinstein 1998).

The puzzle we have used to motivate prospect theory is called the *Allais puzzle* (referring to an article by Allais 1953). In prospect theory, the puzzle is solved by using a single uniform function $\gamma$ that models the deformation of probabilities by certain reference points. However, there are many other puzzles where a similar solution by using such a uniform function $\gamma$ is questionable. In the next section we describe some of these puzzles and make clear that a more systematic approach is desired – one that deals with the origin of probabilities in a cognitive context.

Of course, we do not claim that all problems discussed within the framework of bounded rationality can be solved in a uniform way and each kind of heuristic component is obsolete. However, for a series of puzzles there seems to be a systematic and explanatory way to account for them – one that is quite different from the ideas underlying prospect theory and its follow-ups (e.g., Tversky and Kahneman 1992). The approach we are proposing is exploiting quantum principles following several recent proposals (Aerts 1982, 2009; Blutner 2009; Busemeyer and Bruza 2012; Conte 1983; Franco 2007a; Khrennikov 2010; Yukalov and Sornette 2010).

# 3. More puzzles

### 3.1 The Ellsberg puzzle

In handling the Allais paradox it has been assumed that there are specified probabilities accessible to the decision maker. Prospect theory describes how these probabilities are deformed and a decision is made on the basis of (1b). However, in many real decision situations the probabilities are unknown and not specified in advance (in the case of ignorance). Ellsberg (1961) describes what can happen in such a situation. Assume a ball will be randomly sampled from an urn that contains red (*R*), green (*G*) and blue (*B*) balls and the actor can win a certain payoff if she correctly predicts the colour of the ball. The actor is told that the box contains 300 balls, exactly 100 of them are red and the remaining ones are green or blue but the exact proportion of green or blue balls is not known.  Consider a first pair of choices, between action *R* (voting for red) and *G* (voting for green). Most respondents chose *R*, possibly because they assume the probability that red will be taken is 1/3 but the probability for green is unknown (bounded only in the interval from 0 to 2/3). This example illuminates the distinction between a risky decision under uncertainty and a decision based on ignorance (cf. Russell and Norvig 1995). Consider now the alternative choice between $\bar{R}$ (voting for green or blue) and $\bar{G}$ (voting for red or blue). For a similar reason, most respondents prefer $\bar{R}$ over $\bar{G}$ in this case.

It is easy to see that the pattern {R $\succ$ *G*, $\bar{R} \succ \bar{G}$ } is incompatible with classical expected utility theory. Assume a payoff is $X in case of choosing the correct colour and $0 in the other case. Then in classical utility $R \succ G$ means P(*R*) · u($X) > P(*G*) · u($X), i.e. P(*R*) > P(*G*). On the other hand $\bar{R} \succ \bar{G}$ means P(*G*∨*B*) · u($X) > P(*R*∨*B*) · u($X), i.e. P(*G*∨*B*) > P(*R*∨*B*). Assuming additivity of classical probabilities (for all possible events) we get P(*G*)+P(*B*) > P(*R*) + P(*B*), i.e. P(*G*) > P(*R*). The latter conflicts with the earlier assumption that P(*R*) > P(*G*).

Tversky & Fox (1995) consider a modification of prospect theory in order to resolve the puzzle. Obviously, when the probabilities are unknown, we cannot describe decision weights as a simple transformation $\gamma$ of the probability scale as in (1b). Instead, Tversky & Fox (1995) introduce a weighting function *W* operating directly on the algebra of events.  This weighting function realizes what is called a *capacity* in theories of reasoning with uncertainty (Halpern 2003). In short, it is a normalized function (assigning 0 to the impossible event and 1 to the certain event) that satisfies monotonicity: if *A*⊆*B* then *W*(*A*) ≤ *W*(*B*).  Capacities do not require additivity. I.e., the Kolmogorovian assumption that *W*(*A*∪*B*) = *W*(*A*)+*W*(*B*) for incompatible events *A* and *B* is normally violated. However, a condition called subadditivity is postulated for *W*:

(2) $W(A)+W(B) \leq W(A \cup B)$ = for all incompatible events *A* and *B*.[2]

Using the function *W* instead of a deformed probability function, the new expression for the expected utility of a prospects for the events $E_i$ with outcomes $x_i$ becomes

(3) $U(f) = \Sigma_i W(E_i) \cdot v(x_i)$

Then the scenario {R $\succ$ *G*, $\bar{R} \succ \bar{G}$ } is equivalent to the conditions {$W(R) > W(G)$, $W(G \cup B) > W(R \cup B)$}. If *W* is a subadditive capacity, then the latter condition can be satisfied. For instance, assume $W(R)$ = 1/3, $W(G)$ = 0, $W(G \cup B)$ = 2/3, and $W(R \cup B)$ = 1/3}. Obviously, these stipulations satisfy the two inequalities. Further, it can be seen that these stipulations define a special case of a subadditive capacity which is called an *inner measure* (Halpern 2003).

Of course, this account is not really explanatory since the corresponding assumptions are not motivated independently. Further, one may wonder how ignorance and proper uncertainty differ in principle and the answer is rather unclear. In Section 4, we will come back to this issue.

### 3.2 The disjunction effect

The disjunction effect (Tversky and Shafir 1992) occurs when conditioned decisions are considered. It is closely connected to violations of the 'sure-thing principle', one of the basic claims made by a (classically) rational theory of decision making. In decision making, this principle is just a psychological version of the law of total probability, which we will explain later. Let us assume that a decision maker prefers option *B* over option $\bar{B}$ when knowing that event *A* occurs and also when knowing that event *A* does not occur. Then the 'sure-thing principle' claims that the decision maker should prefer *B* over $\bar{B}$ when not knowing whether or not *A* occurs. If the decision maker refuses *B* (or prefers $\bar{B}$), we have a violation of this principle.

In everyday reasoning, human behaviour is not always consistent with the sure thing principle. For example, Tversky & Shafir (1992) reported that more students would purchase a non-refundable Hawaiian vacation if they were to know that they had passed or failed an important exam, compared to a situation where the exam outcome was unknown. Specifically, $P(B/A)$ = 0.54, $P(B/\bar{A})$ = 0.57, and $P(B)$ = 0.32, where *B* stands for the event of purchasing a Hawaiian vacation, *A* for the event of passing the exam, $\bar{A}$ for the event of not passing the exam, and P for the averaged judgments of probability. Disjunction fallacies are fairly common in behaviour (Busemeyer and Bruza 2012).

Classical probability theory does not allow patterns such as {$P(B/A) > ½$, $P(B/\bar{A}) > ½$ ,$P(B) \leq ½$}. This is a simple consequence from the law of total probability that can be stated as follows:

(4) $P(B) = P(B|A) \cdot P(A) + P(B|\bar{A}) \cdot P(\bar{A})$.

If we assume that $P(B|A) > ½$ and $P(B|\bar{A}) > ½$, then we get P(B) > ½ as a consequence. This conflicts with the assumption $P(B) \leq ½$ comprising the disjunction fallacy.

It is helpful to examine how the law (4) of total probability arises. We require three assumptions. First, we assume that the underlying algebra of events is Boolean. That means essentially, that we have distributivity. In the present case, distributivity allows us to derive

---

[2] The same is required for the dual of *W* defined by the function *W '*(A) = 1- W($\bar{A}$).

$B = BA \cup B\overline{A}$. Second, we assume that probability is an additive measure function. In particular, we have $P(B) = P(BA) + P(B\overline{A})$ for the two disjoint conjunctive events $BA$ and $B\overline{A}$. Third, the standard definition of conditional probability allows us to write:

(5) $P(X|Y) = P(XY)/P(Y)$

The above three assumptions readily derive the law of total probability and explain the classical requirement that there is no disjunction effect, i.e. the difference $P(B) - (P(B|A) \cdot P(A) + P(B|\overline{A}) \cdot P(\overline{A}))$ is always required to be zero.

Unfortunately, prospect theory or its modifications based on the general conception of subadditive capacities is not sufficient to explain the disjunction effect. The reason is that capacities are still based on the idea of a (partial) Boolean algebra and the law of subadditivity is valid within this framework, taking the following form:

(6) $W(B \cap A) + W(B \cap \overline{A}) \leq W(B)$

Using the corresponding modification of (5) for the conditioned weight function $W(B|A) =_{def} W(B \cap A)/W(A)$, it is not difficult to derive $min\{W(B|A), W(B|\overline{A})\} \leq W(B)$. Hence, scenarios such as $\{W(B|A) = 0.54, W(B|\overline{A}) = 0.57, W(B) = 0.32\}$ as in the Hawaiian vacation example are theoretically excluded.

The original explanation Tversky & Shafir (1992) gave for the disjunction effect referred to a psychologically plausible intuition, which corresponds to a failure of plausible reasoning under the unknown condition:

> We attribute this pattern of preference to the loss of acuity induced by the presence of uncertainty. Once the out-come of the exam is known, the student has good- albeit different - reasons for taking the trip: If the student has passed the exam, the vacation is presumably seen as a reward following a successful semester; if the student has failed the exam, the vacation becomes a consolation and time to recuperate before taking the exam again. A student who does not know the outcome of the exam, however, has less clear reasons for going to Hawaii. In particular, she may feel certain about wanting to go if she passes the exam, but unsure about whether she would want to go if she fails. Furthermore, she may feel it is inappropriate to reward herself with a trip to Hawaii regardless of whether she passes or fails. Only when she focuses exclusively on the possibility of passing and of failing the exam does her preference for going to Hawaii become clear. The presence of uncertainty, we suggest, tends to blur the picture and makes it harder for people to see through the implications of each outcome. Broadening the focus of attention can lead to a loss of acuity. (Tversky & Shafir, 1992, p. 306)

Busemeyer and Bruza (2012, p. 267) note that this psychological explanation is quite consistent with an approach based on interference:

> If choice is based on reasons, then the unknown condition has two good reasons. Somehow these two good reasons cancel out to produce no reason at all! This is analogous to wave interference where two waves meet with one wave rising while the other wave is falling so they cancel out. This analogy generates an interest in exploring quantum models. (Busemeyer & Bruza, 2012, p. 267).

In Section 4, we will develop this idea in considerable mathematical detail.

### 3.3 The conjunction fallacy

The conjunction fallacy was described first by Tversky & Kahneman (1983) and then verified by numerous authors (cf. Busemeyer and Bruza 2012). In one of their experiments subjects are presented with a story such as the following one:

> Linda is 31 years old, single, outspoken and very bright. She majored in philosophy. As a student, she was deeply concerned with issues of discrimination and social justice, and also participated in anti-nuclear demonstrations. (Tversky & Kahneman, 1983, p. 298)

After the presentation of the story the subjects are asked to assess the probabilities of several propositions on a numbered scale. We represent the critical propositions only (together with the averaged judgements of the probabilities):

(7) (A)   Linda is active in the feminist movement                              (0.61)
    (B)   Linda is a bank teller.                                                (0.38)
    (A&B) Linda is a bank teller and is active in the feminist movement         (0.51)

Obviously, the results contradict the Kolmogorov axioms of probability theory: The conjunction of two propositions can never get a higher probability than each of the two conjuncts. Closely related results were obtained when asking for ranking the probabilities of the different events or when asking for frequency judgments (cf. the discussion in Busemeyer and Bruza 2012).

Unfortunately, the conjunction fallacy cannot be explained by using any of the theoretical ideas discussed before. Prospect theory accounts for the fact that small probabilities are overweighted and large probabilities are underweighted in decision making. That does not help in the present case since the conjunction fallacy appears both with large and small probabilities. The idea of generalizing standard probabilities to capacities, which was helpful in case of resolving the Ellsberg puzzle, is not successful in the present case either. Capacities satisfy the condition of monotonicity, and monotonicity is violated if we consider the conjunction fallacy: $A\&B \subseteq B$, but we do not have $P(A\&B) \leq P(B)$. Hence, it needs a new heuristic idea in order to resolve the fallacy.

Tversky & Kahneman (1983) propose the judgmental heuristic of *representativeness*. Representativeness refers to estimating the probability of a situation or event by judging how similar it is to a coherent situation or prototype. As noted by Tversky & Kahneman (1983, p. 299) "representativeness depends on both common and distinctive features (Tversky, 1977), it should be enhanced by the addition of shared features." In the case of the Linda example sketched in (7), *A&B* is more representative for Linda's activities than *A* or *B* considered in isolation. Contrary to the standard extensional logic, this makes *A&B* more probable than its constituents.

Unfortunately, there is no formal outline of a theory of representativeness within the context of judgmental heuristic. In the next section we will outline such a theory within the framework of quantum cognition.[3]

---

[3] A closely related phenomenon is the disjunction fallacy illustrated by examples such as "Linda is neither a bank teller nor a feminist" (see Busemeyer and Bruza 2012, p. 126...). The reader should also be referred to the work of Hampton (1988b, 1988a, 2007) investigating related phenomena in the conceptual domain.

## 3.4 Order effects

Survey researchers have demonstrated repeatedly that the same question often produces quite different answers, depending on the question context (for numerous survey examples, see Sudman and Bradburn 1982; Schuman and Presser 1981). To cite just one particularly well-documented example, a group of (North-American) subjects were asked whether "the United States should let Communist reporters come in here and send back to their papers the news as they see it?" The other group was asked whether "a Communist country like Russia should let American newspaper reporters come in and send back to their papers the news as they see it?" Support for free access for the Communist reporters varied sharply depending on whether that question preceded or followed the question on American reporters. The differences are quite dramatic: in a study of 1950, 36% accepted communist reporters when the communist question came first and 73% accepted them when the question came second. When the study was repeated in 1982, the numbers changed to 55% vs. 75%.

    Schumann and Presser (1981) described two kinds of ordering effects, which they called 'consistency' and 'contrast' effects. The example with the case of accepted communist reporters illustrates the *consistency effect*, where in the context of the other question the answer frequencies are assimilated. In the *contrast* case, the differences of the answer frequencies are enlarged. In a more recent article, Moore (2002) reports on the identification of two different types of question-order effects termed as 'additive' and 'subtractive'. The following figure gives a schematic sketch of the four different types of question ordering effects (adapted from Blutner 2012):

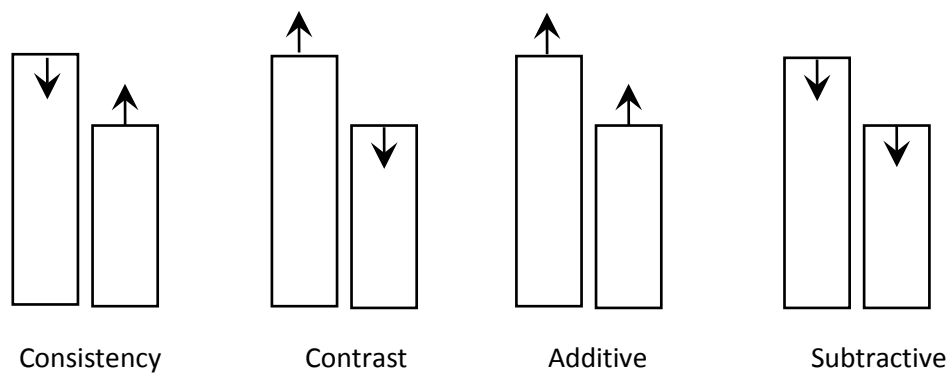

Consistency      Contrast      Additive      Subtractive

Figure 2: Four types of order effects for attitude questions. The size of the blocks for the two questions indicates the percentage of 'yes'-answers when the questions appear isolated. The arrows indicate whether the percentage of 'yes'-answers to these questions increases or decreases if the question is preceded by the alternative question

Question order effects are normally not considered in the context of bounded rationality. The reason is simply that the theoretical models that are standardly deliberated in order to describe the data – such as prospect theory – are actually not used for handling question-ordering effects. Instead, a completely different group of phenomenological models is applied to get insights into the nature of order effects. For instance, Weber and Johnson (Weber and Johnson 2006, 2009; Johnson et al. 2007; Johnson and Busemeyer 2010) have proposed that certain memory processes can be used to model such tasks:

> This approach, most recently dubbed 'Query theory', assumes that preferences that drive choice and other decisions are based on a collection of serially posed queries to memory concerning relevant characteristics of the task. For example, if deciding whether to buy a certain digital camera, an individual might attempt to recall experiences with similar models or generate the pros and cons of buying the camera. Query theory is able to explain some empirical trends in human decision behavior by embellishing this simple notion with what is known about human memory, such as serial position effects, priming, and interference. Although the theory's assumptions have been empirically supported, at this point it has not been formally introduced as a mathematical model or at a specific algorithmic level, as the preceding computational models have." (Johnson & Busemeyer 2010, p. 745).

However, in the theoretical context of quantum models, the situation has changed radically and we will see that the existence of such order effects is constitutive for the appearance of the typical phenomena of bounded rationality. Hence, whether a given phenomenon is meaningfully grouped under a certain label, let it be the label of 'bounded rationality' or any other label, depends on the theoretical paradigm that aims to capture the phenomenon. It is not surprising that prospect theory and quantum cognition generate different groupings. Regarding 'query theory', we can add that it designates a system of informal ideas that possibly can be explicated and mathematized in the formalism of quantum cognition (rather than in terms of prospect theory).

# 4. Quantum probabilities and the proper analysis of some puzzles of bounded rationality

In the last section, we have introduced some puzzles of bounded rationality. We have shown how the Allais paradox can be resolved by prospect theory using a single uniform function γ that models the deformation of probabilities by certain reference points. In contrast, for solving the Ellsberg puzzle a substantial modification of prospect theory was required introducing a novel weighting function. This function realizes what is called a *capacity* in formal theories of reasoning with uncertainty (Halpern 2003). Importantly, the disjunction effect, by contrast, cannot be explained by this modification of prospect theory. Instead, a new idea is required based on the failure of plausible reasoning when the conditions are unknown. Yet another new idea, the concept of representativeness, is required to resolve the conjunction and disjunction fallacies. Some readers might have the impression that the field of judgmental heuristic is dominated by a rather eclectic methodology. We actually agree with this view. Interestingly, some authors suggest to making a virtue out of necessity. For instance, Gigerenzer & Selten (2001) proposed an "adaptive toolbox" to handle the problems. In our opinion, this conclusion is premature and questionable.[4]

In order to avoid misunderstandings, we do not disapprove the empirical findings of judgemental heuristics including the results of Gigerenzer and other representatives of the toolbox idea. Rather, we challenge the fruitfulness of those theoretical ideas for the foundation of cognitive science. In natural sciences, it is not a very attractive methodology to propose a toolbox of different theories in order to handle closely related empirical data. Instead, natural scientists try to explain the data on hand by means of deduction from a universal theory which is the end of their striving. We think this idea should apply also to psychology, linguistics and cognitive science in general. In this section we suggest the formalism of quantum theory as a powerful research instrument that can provide uniform and systematic solutions.

### 4.1 An elementary introduction to Hilbert space semantics

One of the main arguments for a quantum approach to cognitive phenomena is the existence of interference effects in higher cognitive processes such as perception, decision making, and reasoning (Aerts 2009; Blutner 2009; Bruza et al. 2009; Busemeyer and Bruza 2012; Busemeyer et al. 2011; Conte 1983; Conte et al. 2008; Franco 2007a, 2007b; Khrennikov 2006; Pothos and Busemeyer 2009; Primas 2007). In quantum cognition, interference terms typically appear when considering the probability of composed events or propositions.

After a concise introduction of the mathematics of Hilbert spaces, we will explain the basic concepts of quantum probability and the concept of measurement. It goes without saying that we can introduce the mathematical framework in a rather sketchy way here only. Readers who are interested in a deeper mathematical understanding are referred to the rich literature of (text-) books on quantum theory, quantum information, and quantum logic (e.g., Engesser et al. 2009; Pitowsky 1989; Vedral 2006).

---

[4] A forerunner of the toolbox idea is Werner Müller who proposed the project of 'systematic heuristic' (Müller 1970). This project was supported by Walter Ulbricht by a huge research institution (which was, unfortunately, eliminated by Erich Honecker shortly after the fall of Ulbricht).

In traditional theories of formal semantics (Lewis 1983; Montague 1970), the building blocks of the underlying ontology are possible worlds, i.e. unstructured objects without any intrinsic properties and without any relations to other objects. Propositions are considered as sets of possible worlds, and the algebra of propositions is a Boolean algebra (formed by the set-theoretic operations intersection, union, and complement). In contrast, the ontology of quantum theory is based on vectors in a Hilbert space $\mathcal{H}$. Vectors are entities that can be added with other vectors and that can be multiplied with a scalar number. For instance, multiplying a vector $u \in \mathcal{H}$ by the number 1.5 gives a new vector written 1.5 $u$ (which is equivalent to the sum $u+0.5\ u$). It should be noted that vectors can also be multiplied with complex numbers.

Complex numbers are nothing else than pairs $[a_1, a_2]$ of real numbers. The first part of a complex number is called its real part; the second part is called its imaginary part. Complex numbers are usually written in the form $a = a_1 + a_2$ i, with real numbers $a_1$ and $a_2$. For calculating with complex numbers the same rules are used as for calculating with real numbers, respecting the assumption $i^2 = -1$. Complex numbers were introduced to allow for solutions of certain equations, such as $z^2+1 = 0$, that have no real solution (since the square of $z$ is 0 or positive, so $z^2 + 1$ cannot be zero). It is easy to see that the equation has two complex solutions: $z = \pm$ i. Generalizing the result, in 1799 Gauss published the first proof that an $n$th degree algebraic equation has $n$ roots each of the form $a = a_1 + a_2$ i, for some real numbers $a_1$ and $a_2$.

Through the Euler formula, a complex number $a = a_1 + a_2$ i may be written in the form

(8) $\quad a = |a|\ (\cos\theta + \mathrm{i} \sin\theta) = |a|\ \mathrm{e}^{\mathrm{i}\theta}$, where $|a| = \sqrt{a_1^{\,2} + a_2^{\,2}}$ and $\theta = \arctan(\frac{a_2}{a_1})$,

where $|a|$ designates the absolute magnitude of the complex number and $\theta$ is the angle between the abscissa and the vector from the origin pointing to the point $[a_1, a_2]$ in the two-dimensional geometric representation of the Gaussian number plane. Each complex number $a = a_1 + a_2$ i has a complex conjugate $a^* = a_1 - a_2$ i. The product $a\ a^*$ equals $|a|^2 = a_1^{\,2} + a_2^{\,2}$ and provides the *modulus* of the complex number $a$.[5]

Another important Hilbert space operation is the *scalar product*. Scalar multiplication and scalar product are different operations. Whereas the former is an operation between (real or complex) numbers and vectors resulting in vectors; the latter is an operation between two vectors resulting in a (real or complex) number. It is intended to express the similarity of the two vectors. The scalar product of two vectors $u$, $v$ in a given vector space is written in the form u·v. If the scalar product of two vectors is zero, the vectors are to be called 'orthogonal'. The scalar product of a vector with itself defines the square of the 'length' of the vector: $||u||^2 = u \cdot u$.[6]

Vector spaces (denoted by $U$, $V$, …) are sets of vectors that are closed under the two operations of vector addition and scalar multiplication. In other words, if all the vectors $u_i$ (i= 1, …, n) are elements of a vector space, then each *linear combination* of them, i.e. each sum $x_1 \cdot u_1 + x_2 \cdot u_2 + ... + x_n \cdot u_n$, is also an element of the vector space. Vector spaces based on scalar multiplication with real numbers are called *real vector spaces*; vector spaces based on scalar multiplication with complex numbers are called *complex vector spaces*.

---

[5] The usefulness of complex numbers becomes evident when we consider vectors that represent oscillations (the appearance of waves at a certain point of space). Such oscillations can be represented by functions of time such as $\mathrm{e}^{\mathrm{i}2\pi f t}$ (where $f$ is the frequency of the oscillation). The scalar multiplication of this 'vector' with the complex number $\mathrm{e}^{\mathrm{i}\theta}$ results in a phase shift of the original oscillation: $\mathrm{e}^{\mathrm{i}\theta}\ \mathrm{e}^{\mathrm{i}2\pi f t} = \mathrm{e}^{\mathrm{i}(2\pi f t + \theta)}$.

[6] Mathematically spoken, the scalar product is a sesquilinear form, i.e. an antilinear form in the first argument and a linear form in the second argument: $au \cdot bv = a^*b\ u \cdot v$; $(u + u') \cdot (v+v') = u \cdot v + u' \cdot v + u \cdot v' + u' \cdot v'$.

A subset of a vector space $U$ is called a *subspace* of $U$ if it is a vector space in itself (i.e. closed under addition and scalar multiplication). Let $U_1$ and $U_2$ be two subspaces of $U$, then the *sum* of the two vector spaces, written $U_1 + U_2$, is the set of all possible sums of elements of $U_1$ and $U_2$. The sum of two vector spaces is a vector space again. We will say that the sum $U_1 + U_2$ is a *direct sum*, written $U_1 \oplus U_2$, of the two subspaces $U_1$ and $U_2$ if and only if each element of the sum can be written uniquely as a sum $u_1 + u_2$ where $u_1 \in U_1$ and $u_2 \in U_2$. It can be proven (e.g. Axler 1996) that a sum $U_1 + U_2$ is a direct sum if and only if $U_1 \cap U_2 = \{0\}$. The *linear hull* of a list of vectors $(u_1, u_2, ..., u_n)$ in $U$ is defined as the set of all linear combinations of these vectors, denoted $LH(u_1, u_2, ..., u_n)$. A vector space is called *finite-dimensional* if it is the linear hull of some finite list of vectors.

An important concept is linear independence of a set of vectors. A set of vectors is called *linearly independent* if none of its elements is a linear combination of the others. Otherwise it is called *linearly dependent*. A *basis* of a vector space $U$ is a list of linearly independent vectors in $U$ iff $U$ is the linear hull of these vectors. If there are several bases of a vector space, it can be proven that the number of the vectors is the same in each base. This number is called the *dimension* of the vector space. Observe that $dim(U_1 + U_2) = dim(U_1) + dim(U_2) - dim(U_1 \cap U_2)$. This result allows the following conclusion: the dimension of the *direct sum* of two vector spaces is the sum of the dimensions of the two spaces. A basis of a vector space can consist of vectors which are pairwise orthogonal and each has unit length. Such a basis is called an orthonormal basis. Is it easy to show that the scalar product of two vectors is the sum of the products of their respective coefficients taken an orthonormal basis.

Let us consider a subspace $U$ of a Hilbert space $\mathcal{H}$. With the help of the scalar product the orthocomplement of $U$ – written $U^\perp$ – can be defined as the set of vectors that are orthogonal to each vector in $U$: $U^\perp = \{u \in \mathcal{H}: u \cdot v = 0 \text{ for any } v \in U\}$. It is not difficult to prove that the orthocomplement is a vector space again and that $\mathcal{H}$ is the direct sum of $U$ and $U^\perp$: $\mathcal{H} = U \oplus U^\perp$. The algebra that arises from considering the three basic operations on vector spaces, intersection, sum, and orthocomplement is called an orthomodular lattice and will be considered more closely in Section 6. At this moment it is sufficient to say that this algebra has properties similar to a Boolean algebra with the exception that the principle of distributivity is lacking. Considering all subspaces of a Hilbert space $\mathcal{H}$ is useful then when it comes to look for alternatives to the classical conception of propositions (sets of possible worlds).

Closely related to subspaces of a Hilbert space $\mathcal{H}$ are so-called projection operators **A**. They project each vector of $\mathcal{H}$ into a given subspace of $\mathcal{H}$.[7] Formally, they can be represented by linear operators, i.e. operators that satisfy the condition **F**$(au + bv) = a$ **F**$(u) + b$ **F**$(v)$ for any complex numbers $a$ and $b$ and any vectors $u, v \in \mathcal{H}$. They have to satisfy the special property of idempotence: **AA**=**A**. Equivalently, this can be expressed by the conditions that projection operators have eigenvalues 0 or 1.[8] In the quantum approach, propositions are modelled by projection operators.

---

[7] That means, if we decompose any vector $u$ into the direct sum $u_1 \oplus u_2$, where $u_1$ is in the given subspace and $u_2$ in its orthocomplement, then **A**$(u) = u_1$.

[8] The eigenvectors of a linear operator **F** are the vectors u satisfying the equation **F**$(u) = \lambda\, u$ for a (complex) number $\lambda$. This number is called an eigenvalue. Not all linear operators have eigenvectors. In the following we will consider so-called Hermitean operators. They are defined as operators which always have real eigenvalues. Alternatively, Hermitean operators can be defined by the condition that **F**\* = **F**, where **F**\* denotes

This is equivalent of modelling them by subspaces of $\mathcal{H}$. The subspace can be reconstructed as the positive eigenspace of **A**, i.e. the set of all vectors satisfying the eigenvalue equation **A**(*u*) = *u*. In combining projection operators, order can matter. That is, it can be that **AB**≠**BA** for two projection operators **A** and **B**. Interestingly, if all projection operators relative to a given Hilbert space commute (i.e., **AB**=**BA**), then we get a Boolean algebra of projectors. The important conclusion is that the algebra of projection operators contains a Boolean algebra as a special case.

It is instructive to consider the two main differences between the treatment of classical propositions (sets of possible worlds) and quantum propositions (projections in Hilbert spaces). First, instead of union *A*∪*B* in the classical case, we consider the sum operation **A**+**B** in the quantum case (constructing the smallest subspace that contains the two subspaces corresponding to **A** and **B**. The second difference refers to complementation. In the quantum case, a negated proposition refers to a subspace orthogonal to the original one. We will write $\overline{A}$=**I**–**A** for the orthogonal projection operator[9] (**I** is the identity operator mapping any vector to itself). It is easy to see that $A\overline{A} = \overline{A}A$ =(**I**–**A**)**A**=**IA**–**AA**=**A**–**A**=0.

Let us consider now the mathematical notion of probability. Even in the quantum case, a probability function is an additive measure function: P(**A**+**B**) = P(**A**) + P(**B**) assuming that the two considered projectors **A** and **B** are orthogonal to each other. In both the classical and the quantum case, the second condition is normalization, assigning the value 1 to the weakest proposition and the value 0 to the strongest one: P(**I**) = 1 and P(∅) = 0.[10] For each normalized vector *u*, it is possible to define a probability function in the following way:

(9)  P$_u$(**A**) = ||**A**(*u*)||$^2$

This is the so-called Born-rule stating that the square of the length of the projection **A**(*u*) should be seen as probability for the proposition *A*, or, in a physical jargon, the probability that the state *u* collapses into the positive eigenspace of **A** (with eigenvalue 1). It is important to see that this probability function is defined by the geometry of the Hilbert space, especially the geometric features of the projector **A** and the (pure) state *u*. For that reason we will call the probability function P$_u$ a *structural* or *geometric* probability. It contrasts with another notion of probability where a system of orthonormal vectors $u_i$ is given together with (statistical) weights $p_i$ ($\Sigma p_i$ = 1) and the resulting statistical probability is calculated in the following way:

(10)  $P_{\{u_i, p_i\}}(A) = \sum_i p_i \cdot P_{u_i}(A)$

---

the adjoint of **F**. Hereby the *adjoint* is defined by the following clause: (**F**v)·*u* = v·(**F***u*) for all vectors *u*, v ∈ $\mathcal{H}$. Projection operators are always Hermitean operators.

[9] Sometimes the symbol **A**$^\perp$ is used for indicating the orthocomplement. Since, from the context, it is always clear whether orthocomplementation is meant or the usual set-theoretic complement, we will use the form $\overline{A}$ in both cases.

[10] It should be stressed that we are using a subjective (or Bayesian) notion of probability in the present context. In this regards, it is essential that there is a decision theoretic foundation of the additive measure function both in the set theoretic case and in the case of considering the projection lattice of a Hilbert space. As shown in several papers by Fuchs and colleagues (e.g., Barnum et al. 2000; Caves et al. 2002a), there is a standard Dutch book argument in both cases. The Dutch-book argument shows that, to avoid sure loss, an agent's gambling commitments should obey the usual axioms establishing additive measure functions.

This *statistical probability* is relative to the (statistical) weights $p_i$ and can be seen as a weighted mixture of (structural) probabilities. John von Neumann has introduced so-called density matrices in order to express such mixed states in an elegant and uniform way (von Neumann 1932). However, for the present purpose it is sufficient to use the concept of mixture in an informal way.

Next, we have to explain the idea of a physical measurement, where the observable being measured is represented by a Hermitian operator **F**. Assume that the considered physical system is in a certain state $u \in \mathcal{H}$. There are two possibilities now. First, the state *u* is an eigenstate of **F**, say with eigenvalue $\lambda$. Then measurement of **F** yields this eigenvalue and the state after measurement is not changed (i.e. it is *u* again). The second possibility is that the state *u* is not an eigenstate of **F**. In this case, quantum mechanics assumes that the act of measurement changes this state into another state which is always an eigenstate of **F**. However, the particular eigenstate out of several possible ones that becomes actualized is decided by chance. There is no way to formulate a deterministic mechanism for this decision. Hence, indeterminism is an essential component of quantum mechanics. The only thing that can be predicted is the *probability* of finding the output state in a certain eigenspace of **F**. For example, consider the space described by the projection operator $A_i$. The probability that the state *u collapses* to the eigenspace of $A_i$ can be calculated by the Born-rule: $P_u(A_i) = ||A_i(u)||^2$. It yields the probability that the eigenvalue $\lambda_i$ is measured.

To repeat the deep insight from quantum theory: the act of measurement can change the state of the system. Only if the initial state is an eigenstate of the observable being measured the final state is not changed by the measurement. In all other cases, the initial state *u* is changed into a mixed state describing the possible outcomes of the measurement and their probabilities. Generally, a measurement can be seen as a question addressed to nature. The act of questioning can change the state of the system. This is not really surprising when considering modern versions of update semantics (Blutner 2012).

If an exact outcome of the measurement cannot be calculated, what is the expectation value of measuring observable **F** when measurement takes place in a pure state *u*? Based on the intuitive idea of measurements, the scalar product between *u* and the outcome of applying the observable **F** to the state *u* can be used to calculate this expectation value. This is labelled by $F_u$.

(11)  $F_u = u \cdot F(u)$

In case of projection operators, formula (11) gives the probability of the proposition expressed by the projector: $A_u = u \cdot A(u) = u \cdot A(A(u)) = A(u) \cdot A(u) = ||A(u)||^2 = P_u(A)$.

Lüders (1951) has proposed a formula that refers to a sequence of two measurements *A* and then *B*. The corresponding operator we will abbreviate (***A***; ***B***). According to Lüders the probability for such a sequence can be calculated as follows:

(12)  $P_u(A; B) = ||B(A(u))||^2$

Hence, the operator **A** is first applied to the state u, and then the operator **B** is applied to the resulting state and transforms it into a final state whose squared length determines the structural probability of the sequence.

In order to determine the operator representing the sequence $(A; B)$ we apply the Born rule (9) to the state $v = A(u)$, i.e. $P_v(B) = ||B(v)||^2 = B(v) \cdot B(v) = B(A(u)) \cdot B(A(u)) = BA(u) \cdot BA(u) = A(u) \cdot BBA(u) = u \cdot ABA(u)$. Therefore the following equation holds (Niestegge 2008):

(13) (**A**; **B**) = **ABA**

This sequence operator can be used to define conditionalized probabilities in the quantum case (Niestegge, 2008):

(14) P(**B**|**A**) = P(**ABA**)/P(**A**) .

The operator (**A**; **B**) or **ABA** is called asymmetric conjunction. Note that P(**ABA**) = P(**A** and then **B**), which is how Busemeyer et al. (2011) modelled conjunction in human decision making (see also Blutner, 2009).[11] Note further that standard systems of quantum logic (Engesser et al. 2009) always use symmetric versions of conjunction and have deep intrinsic problems when it comes to consider non-commuting projections. One problem is that the resulting systems do not obey the deduction theorem which is constitutive for a proper logic in the opinion of many logicians. In the following, we will see that the notion of asymmetric conjunction is very useful when discussing certain puzzles of bounded rationality.

### 4.2 The Allais puzzle

In Section 2 we have considered four prospects *a,b,c,d*, representing pairs of one or two winning situations (to win 1000 units or to win 450 units) and one losing situation (to win nothing). In the classical case, we can represent the utilities as the diagonal elements of the following matrix:

$$(15) \quad U = \begin{pmatrix} u(1000) & 0 & 0 \\ 0 & u(450) & 0 \\ 0 & 0 & 0 \end{pmatrix}$$

The four prospects are then defined by the vectors $a = (\sqrt{.5}, 0, \sqrt{.5})$, $b = (0, 1, 0)$, $c = (\sqrt{.1}, 0, \sqrt{.9})$, $d = (0, \sqrt{.2}, \sqrt{.8})$. The single components of these vectors give the square root of the probability of the corresponding situation. Let us call such vectors risk profiles. The expected utility of any risk profile *x* can then be calculated by the quadratic form:[12]

(16) $U_x = x \cdot U x^T$

It is easy to check that this formula gives exactly the results we found earlier for the expected utility in the classical model: $U_a$ = .5 *u*(1000), $U_b$ = *u*(450), $U_c$ = .1 *u*(1000), and $U_d$ = .2 *u*(450). In this way, it is not possible to reproduce the experimentally expected results that prospect *b* is preferred to *a* while *c* is preferred to *d*. For example, if we assume that u(1000) = 10 and u(450) = 5.5, then we get correctly that *b* is preferred to *a* but wrongly that *d* is preferred to *c*.

---

[11] For a different but related approach see Yukalov and Sornette (2010, 2013).

[12] In general, $F_x = x \cdot F x^T$ can be seen as matrix pendant to formula (11) expressing the expectation value in the abstract operator formalism. In the matrix calculations, the application operation corresponds to the product of a matrix with a (column) vector, and the scalar multiplication corresponds to the multiplication of a row vector with a column vector (resulting in the sum of the products of the corresponding component values).

La Mura's idea of 'projected expected utility' (La Mura 2009) is to consider formula (16) as valid even when $U$ is not a diagonal matrix but contains non-diagonal elements different from zero. For example, he suggests to assuming a slight aversion to the risk of obtaining no gain (third component of the vectors) such that we have to consider matrices as the following one:

$$(17)\ U = \begin{pmatrix} 10 & 0 & -1 \\ 0 & 5.5 & -1 \\ -1 & -1 & 0 \end{pmatrix}$$

Using formula (16) again now yields the following results: $U_a$ = 4, $U_b$ = 5.5, $U_c$ = .4, and $U_d$ = .3. Hence, prospect *b* is preferred to *a* while *c* is preferred to *d*.

### 4.3 The Ellsberg puzzle

La Mura's theory of 'projected expected utility' can be applied to resolving the Ellsberg puzzle as well (La Mura 2009). In Section 3.1 we have considered two prospects *R* and *G* assuming a payoff of $X in case of choosing the correct colour and $0 in the other case. Hence, instead of two winning situations as in the example discussed before, we are concerned with one winning situation only (to win $X) and one losing situation (zero-payoff). For simplicity, we assume that u($X) = 1 and U($0) = 0. Hence, we can model the scenario by assuming the following diagonal utility matrix:

$$(18)\ U = \begin{pmatrix} 1 & 0 \\ 0 & 0 \end{pmatrix}$$

The two prospects R and G can modelled then by two risk profiles: $r = \left(\sqrt{\frac{1}{3}}, \sqrt{\frac{2}{3}}\right)$ and $g = \left(\sqrt{\frac{j}{300}}, \sqrt{\frac{300-j}{300}}\right)_{0 \leq j \leq 200}$. In the latter case the risk profile contains a hidden parameter *j* ranging from 0 to 200 (the number of green balls the actor does not know). We will call such risk profiles mixed risk profiles. It is easy to calculate the expected utilities for both profiles: $U_r$ = 1/3 and $U_g$ = j/300. In the latter case we will assume that all possibilities for choosing *j* are equally mixed. In the average, then we get $mean\ U_g = \frac{1}{201}\sum_{j=0}^{200}(\frac{j}{300})$ = 1/3. This is the classical solution, which does not make a difference between risk and ignorance. Consequently, both prospects are ranked equally.

Again, we consider formula (16) as valid even when *U* is not a diagonal matrix but contains non-diagonal elements different from zero. Following La Mura (2009), we consider the following Hermitean utility matrix containing a free parameter $\alpha$ (real number):

$$(19)\ U = \begin{pmatrix} 1 & \alpha \\ \alpha & 0 \end{pmatrix}.$$

Again, we can calculate the expected utilities and obtain now: $U_r$ = 1/3 + 0.94$\alpha$ and $U_g$ = j/300 + $2\alpha\sqrt{j(300-j)}$/300. Assuming again that all possibilities for choosing *j* are equally mixed, we get *mean* U$_g$=$\frac{1}{201}\sum_{j=0}^{200}(\frac{j+2\alpha\sqrt{j(300-j)}}{300})$ = 1/3+0.83α.[13] If α = 0, there is no difference in preference between the risky action R and the uncertain action (action of ignorance) G. If α > 0, then the risky action *R* is advantageous over the uncertain action *G* (and converse in the case of α < 0). Hence, the empirical data can be fitted by a positive parameter α determining the avoidance of situations of ignorance.

Summarizing, the probability of risky events is calculated by the Born rule; the probability of events under ignorance is calculated by mixing. Quantum probabilities correspond to the normal case of judging risks. The case of ignorance (no explicit probabilities are provided by the geometry of projections) are handled by the mixed case of the density matrix (Franco 2007c).[14]

### 4.4 The disjunction effect

In Section 3.2 we have seen that classical probability theory cannot account for the disjunction effect. The reason was that in the classical theory we can derive the law of total probability (4), for convenience repeated here:

(20)   P(*B*) = P(*B*|*A*)·P(*A*) + P(*B*|$\overline{A}$)·P($\overline{A}$).

Let us see now what happens with equation (20) in the quantum case. As mentioned above, also in the quantum case, a probability function is an additive measure function. However, instead of using sets of possible worlds to model propositions, the quantum approach models propositions by subspaces of a given Hilbert space, or projection operators that project any vector into the given subspace. In order to get the quantum version of equation (20), we can decompose the projector ***B*** in the following way:

(21)   ***B***=***IBI*** $= (A + \overline{A})B(A + \overline{A})$ =***ABA***+$\overline{A}B\overline{A}$+***AB***$\overline{A}$+$\overline{A}$***BA***

The four parts of this decomposition are orthogonal to each other. Using the definition of asymmetric conjunction as given in equation (13), we get (22a); using in addition the definition of conditionalized probabilities – equation (14) –, we obtain (22b).

(22)   a. P(***B***) =   P(***A; B***)   + P($\overline{A}$; ***B***)   + ∂(***A***,***B***), where ∂(***A***,***B***)=P(***AB***$\overline{A}$+$\overline{A}$***BA***)
        b. P(***B***) = P(***B/A***) P(***A***) + P(***B***|$\overline{A}$) P($\overline{A}$) + ∂(***A***,***B***),

---

[13] Instead of considering the sum $\frac{1}{201}\sum_{j=0}^{200}(\frac{j+2\alpha\sqrt{j(300-j)}}{300})$, of course, it is possible to consider the definite integral $\frac{1}{201}\int_{0}^{200}\frac{j+2\alpha\sqrt{j(300-j)}}{300}dj$. Busemeyer and Bruza (2012, p. 260) who consider three possibilities only (all blue balls, all green balls, equal number of blue and green balls) gets a mean of 1/3+0.63α instead of 1/3+0.83α, and La Mura (2009) who considers two possibilities only gets 1/3+0.47α.

[14] In standard treatments of reasoning with uncertainty, the difference between risk and uncertainty is treated by Dempster-Shafer theory (Russell and Norvig 1995) In this theory, the introduction of ignorance leads to the specification of intervals. The bigger the span of the interval the bigger the degree of ignorance. A careful comparison between the Dempster-Shafer approach and the present account goes beyond the scope of this paper.

The term $\partial(\boldsymbol{A},\boldsymbol{B})$ is called the interference term. It is zero if **A** and **B** commute, in which case equation (22b) reduces to (20) – the law of total probability.

Equation (22b) allows to expressing the disjunction effect. In fact, the difference $P(B) - (P(B|A) \cdot P(A) + P(B|\overline{A}) \cdot P(\overline{A}))$ is the interference term $\partial(\boldsymbol{A},\boldsymbol{B}) = P(AB\overline{A}+\overline{A}BA)$. Hence, we get a numerical value for the disjunction effect when we calculate the interference term. In quantum theory, the probability function P is always relative to the state of the system, either mixed or pure. For simplicity, we consider pure states $u$ only at the moment. In this case, we can calculate the following expression for the interference term:

$$(23) \quad P_u(AB\overline{A}+\overline{A}BA) = 2\sqrt{P_u(B|A)P_u(A)} \cdot \sqrt{P_u(B|\overline{A})P_u(\overline{A})} \cdot \cos(\Delta).\ ^{15}$$

The phase shift parameter $\Delta$ relates to the impact of knowing $A$ or $\overline{A}$ for assessing the likelihood of $B$. This angle is zero if the subspaces corresponding to the events $A$ and then $B$ (or $\overline{A}$ and then $B$) are orthogonal. If they are not orthogonal, the subspaces are incompatible. This means that if a participant decides for $B$, then he/she must necessarily be undecided with regard to $A$. From a psychological perspective, the interference term is the correlation between two decision paths: For the Hawaiian vacation example considered in Section 3.2, (i) First consider you won't pass the exam and then consider the trip to Hawaii and (ii) first consider you will pass the exam and then consider the trip to Hawaii. A negative correlation corresponds to a negative interference term ($\partial(\boldsymbol{A},\boldsymbol{B})<0$ in (22)), which negatively affects the law of total probability (i.e., reduces the probability for the trip, in the unknown case), and conversely for a positive correlation.

Considering the numerical values of the Hawaiian vacation example, $\{P(B|A) = 0.54, P(B|\overline{A}) = 0.57, P(B) = 0.32\}$, we get a value of -.23 for the interference term (assuming the chances for passing and not passing are equal), and from this outcome we can fit the phase shift parameter: $\cos(\Delta) = -0.42$, i.e. $\Delta = 114°$.

### 4.5 The conjunction fallacy

We will demonstrate now how asymmetric conjunction resolves this conjunction fallacy discussed in Section 3.3 in the quantum probabilistic case. We can define the conjunction effect as difference $P(A;B) - P(B)$ with the actual numerical value of +0.13 found for the Linda-example reviewed above.

Obviously, the results contradict the Kolmogorov axioms of probability theory: The conjunction of two propositions can never get a higher probability than each of the two conjuncts. We will demonstrate now how asymmetric conjunction resolves this conjunction puzzle in the quantum probabilistic case. Considering equations (22b) and (23), we get the following expression for the conjunction effect:

$$(24) \quad P(A;B) - P(B) = - P(\overline{A};B) - 2 \cdot \sqrt{P(A;B) \cdot P(\overline{A};B)} \cdot \cos(\Delta).$$

According to classical probability theory, the value of the conjunction effect is always negative. This corresponds to the case of commuting operators $A$ and $B$, and can be expressed by assuming $\Delta=\pi/2$ or cos(Δ) = 0). The conjunction effect can be positive if cos(Δ) is negative (e.g., if cos(Δ)= −1 or Δ=π). In the example case with P(*B*) = 0.38 (Linda is a bank teller), P(*A*) = 0.61 (Linda is a feminist), P(*A*; *B*) = 0.51 (Linda is a feminist bank teller), we get a conjunction effect P(*A*; *B*) − P(*B*) = 0.13. This corresponds to a parameter cos(Δ) = −0.9 assuming P($\overline{A}$; *B*) = 0.09 (Linda is a non-feminist bank teller).

---

[15] Proof of this equation: $P_u(BA\overline{B} + \overline{B}AB) = u \cdot (BA\overline{B} + \overline{B}AB)(u) = (u \cdot \overline{B}AB(u))^* + u \cdot \overline{B}AB(u) = 2\,\mathrm{RE}(u \cdot \overline{B}AB(u)) = 2\,\mathrm{RE}(A\overline{B}(u) \cdot AB(u)) = 2\,||\,A\overline{B}(u)\,||\,\,||\,AB(u)\,||\cdot \cos(\Delta)$.

The present approach followed Conte and Khrennikov (e.g. Conte et al. 2008; Khrennikov 2003a, 2003b, 2006) who pioneered the investigation of interference effects in cognitive macro-systems. The same idea can be used for modelling data on the conjunction and disjunction of concepts. Research pioneered by Hampton and others concerning the structure of vague concepts provides significant empirical basis showing deviations from set theoretic rules in conceptual combination (Hampton 1987, 1988a, 1988b; Budescu and Wallsten 1995). The relevant empirical results include violations of the conjunction and disjunction rules, the famous 'guppy effect', and cases of 'dominance', 'over- and underextension', which were all successfully described on the basis of quantum principles (Aerts 2009; Aerts and Gabora 2005).

For linguists, the distinction between vagueness and prototypicality is very important (Kamp and Partee 1997). While classical models have huge problems to model this distinction, it is quite easy to model it in quantum approaches. Concepts are modelled like propositions as projection operators where the Hilbert space defines the set of instances (feature vectors). The corresponding measure functions define the graded membership function for the described concept. In addition, within the positive eigenspace of this operator there is a low-dimensional subspace defined which describes the relevant prototypes of the category. Projecting into this subspace provides a measure function for typicality (Blutner 2009; Blutner et al. 2013).

### 4.6 Order effects

Recently, Wang and Busemeyer (2013) have shown that the quantum approach can describe all four types of order effects we have considered in Section 3.4. They treat the ordering effect very similar to the disjunction effect, namely in terms of the asymmetric conjunction (see Section 4.4). Let the question order effect be the difference between the probabilities for ***B*** in the contexts ***A*** and $\overline{A}$ and the probability for ***B*** in a neutral context, i.e. the difference P(***A*;** ***B***) + P($\overline{A}$; ***B***) − P(***B***). This difference has the negative amount of the interference term that was introduced in (22): −∂(***A***,***B***) = −P(***AB$\overline{A}$***+$\overline{A}$***BA***). For simplicity, we consider pure states *u* only at the moment as we did in case of calculating the disjunction effect (Section 4.4). Then we obtain the following expression for the question order effect:

(25)  −∂$_u$(***A***, ***B***) = 2P$_u$(***A*;** ***B***) − 2(P$_u$(***A***) P$_u$(***B***))$^{½}$ cosδ

Here, δ is a phase shift parameter introduced by factorizing the complex number defined by the scalar product ***A****u* · ***B****u*.[16] In the classical case of commuting operators the order effect becomes zero.

---

[16] Assuming that we follow the Born-rule for calculating probabilities, we can derive the question ordering effect (25) in the following way:

$$-\partial_u(\boldsymbol{A}, \boldsymbol{B}) = P_u(\boldsymbol{A}; \boldsymbol{B}) + P_u(\overline{A}; \boldsymbol{B}) - P_u(\boldsymbol{B}) =$$
$$\boldsymbol{BA}u \cdot \boldsymbol{BA}u + \boldsymbol{B}\overline{A}u \cdot \boldsymbol{B}\overline{A}u - (\boldsymbol{BA}u + \boldsymbol{B}\overline{A}u) \cdot (\boldsymbol{BA}u + \boldsymbol{B}\overline{A}u) =$$
$$-\boldsymbol{BA}u \cdot \boldsymbol{B}\overline{A}u - \boldsymbol{B}\overline{A}u \cdot \boldsymbol{BA}u =$$
$$-2\,\text{Re}(\boldsymbol{BA}u \cdot \boldsymbol{B}\overline{A}u) =$$

In the non-classical case of non-commuting operators we can describe all four classes of ordering effects. For instance, let us assume that $P_u(A) > P_u(B)$, and $P_u(A; B) > P_u(B; A)$, then the consistency (assimilation) effect is obtained by stipulating $P_u(A; B) > \cos\delta > P_u(B; A)$. Similarly, for the contrast effect we assume $P_u(A; B) < \cos\delta < P_u(B; A)$; and correspondingly for addition and subtraction (cf. Blutner 2012).

The phenomenon of order effects convincingly illustrates that quantum models of cognition are much more powerful than simply fitting parameters to a collection of isolated phenomena. We think that quantum models sometimes have the potential of providing real explanations. This suggests that – once the relevant parameters are fixed – the theory can make predictions about correlations between different effects (Atmanspacher and Römer 2012). An excellent example is provided by Wang's and Busemeyer's (2013) verification of what they call the 'law of reciprocity':[17]

(26) $P(A; B) + P(\overline{A}; B) = P(B; A) + P(\overline{B}; A)$

The theorem can be tested by considering two questions that are answered one immediately after the other with no possibility to insert additional information in between. The yes-answer for a question can be seen as realizing proposition *A* (or *B*, respectively), whereas the no-answer corresponds to the proposition $\overline{A}$ (or $\overline{B}$, respectively). The empirical test of reciprocity (Wang and Busemeyer 2013; see also Busemeyer and Bruza, 2012) was surprisingly successful with one exception, the test of examples producing subtractive effects. It could be argued that in the latter case the key assumption of the model, no intervening information between the two events, is violated.

There is a variety of other examples where quantum models have proven their explanatory value: the simultaneous explanation of the conjunction and disjunction fallacies (Busemeyer & Bruza 2012, p. 126 ff), the prediction of borderline contradictions in case of conjoining vague predicates (Blutner et al. 2013), the prediction and empirical verification of inequalities for complementary questions in the context of Jung's personality theory (Blutner and Hochnadel 2010), and belief revision in update semantics (beim Graben 2014).

### 4.7 General discussion

In Section 3 we have discussed some puzzles of bounded rationality. Classical decision theory based on standard probabilities, i.e. measure functions on the basis of Boolean algebras, cannot resolve these problems and for that reason they have been called puzzles. Further, we have introduced prospect theory and modifications thereof, and we have discussed how these models can help to solve the puzzles. For instance, we have argued that the Allais paradox can be resolved by using a single uniform function $\gamma$ that models the deformation of probabilities by certain reference points, a basic assumption of prospect theory. Further, we have reported that for solving the Ellsberg puzzle a substantial modification of prospect theory is required, one that introduces a new weighting function

---

$$-2\,\mathrm{Re}(\boldsymbol{BA}u \cdot (\boldsymbol{B}-\boldsymbol{BA})u) =$$
$$2\,\boldsymbol{BA}u \cdot \boldsymbol{BA}u - 2\,\mathrm{Re}(\boldsymbol{A}u \cdot \boldsymbol{B}u) =$$
$$2P_u(A; B) - 2(P_u(A)\,P_u(B))^{½} \cos\delta.$$

[17] This theorem is the same as Niestegge's (2008) Axiom 1. Busemeyer and Bruza (2012) have given a proof for the theorem, both by considering pure states and by considering the more general density matrixes.

(called a *capacity* in formal theories of reasoning). Next, we have argued that the disjunction effect cannot be explained by this modification and a new psychological idea is required (based on the failure of plausible reasoning when the conditions are unknown). For resolving conjunction and disjunction fallacies yet another new idea is required – the idea of representativeness.

Summarizing, the impression is unavoidable that prospect theory (and their modifications) make isolated stipulations in order to describe apparently isolated phenomena. This is not really a satisfying theoretical situation (even if some authors have tried to make a virtue out of necessity). Models based on quantum probabilities (i.e., measure functions on the basis of orthomodular lattices such as Hilbert space projection lattices) improve the situation drastically. Allais paradox and Ellsberg puzzle can be resolved by the standard matrix formalism of expected utility where some non-diagonal elements of the utility matrix have to be taken into account. The remaining puzzles are described as interference phenomena and crucially rely on the operation of asymmetric conjunction. It should be stressed that the available models of quantum cognition are still based on several stipulations (size and sign of the interference effect, additional parameters for the utility matrices). With a proper fit of the respective parameters, we are able to describe the available data. However, descriptions are not yet explanations.  Proper explanations are very close to predictions of effects based on independently motivated assumptions. We think that quantum models sometimes have the potential of providing explanations and do more than just describing a list of (isolated) phenomena. Once the relevant parameters are fixed, the theory can make predictions about correlations between different effects (as discussed in Section 4.6).

There is another opportunity to overcome the shortcomings of purely descriptive models. This is the exploration of foundational research questions. In the following sections, we will leave the descriptional level and advance to a more foundational level. This may help to achieve better understanding why the quantum models can be successful.

## 5. Phenomenological and foundational research programs

In the last section, we have outlined how quantum cognition can be seen as a phenomenological research program[18] that solves several puzzles of bounded rationality. Not every cognitive scientist will be convinced by the proposed analyses and some researchers might ask for a deeper motivation of the technical instrument of quantum probabilities. In other words, they might ask *why* this formalism is appropriate. Is there a handful of independently motivated principles that account for the character of quantum probability? In Section 6 we will give a positive answer to this question. This justifies the claim that we are following a research program that can truly be called foundational.

The distinction between phenomenological and foundational research programs is not a mainstream issue in the current philosophy of science. However, it is lively discussed within particular disciplines such as physics (e.g., Streater and Wightman 1964; Piron 1976), chemistry (e.g., Carbó 1995), and linguistics (e.g., Hinzen 2000). With regard to the foundational perspective, we have to distinguish two different kinds of foundational research. One kind intends to reduce phenomenological aspects to a deeper level of description (such as explaining chemical properties by molecular structures). The other type of foundational research tries to get a deeper understanding of the theoretical instruments by reducing them to a small set of independently motivated first principles (Primas 1990). These principles are all stated at the same 'level' of description by a series of axioms or postulates. Accordingly, the second type of foundational research is anti-reductionist.

---

[18] The term 'phenomenological' as used in this paper is not intended to refer to the use of this term in the philosophy of Husserl and Heidegger. In the Husserlian tradition, the phenomenological researcher aims to describe and to understand the meaning of the participants' lived experiences. In the present broader use of this term, we simply refer to any research that tries to analyse/model/explain some empirical phenomenon in the domain under discussion (independently of the feelings, motivations and seeks for 'meanings' of the involved observers)

In the following, we will concentrate on the second way of understanding foundational research. We will give first some examples in order to illustrate several aspects of foundational research and finally we will present some basic traits generally characterizing foundational research in the domain of cognitive science. This paves the way for a proper understanding of the foundational approach to quantum probabilities as considered in later sections.

The first example concerns quantum theory and quantum field theory. In their influential book, Streater and Wightman (1964) stress the needs of an axiomatic treatment of quantum field theory in order to answer such questions as "what is a quantized field?" and "what are the physically indispensable attributes of a quantized field?" Each theoretical achievement requires abstraction and idealizations (Stokhof and van Lambalgen 2011). Without this, a powerful and simple theory is not possible. However, idealizations or abstractions can lead to inconsistences of the whole theoretical system.

> "In fact, the main problem of quantum field theory turned out to be to kill it or cure it: either to show that the idealizations involved in the fundamental notions of the theory (relativistic invariance, quantum mechanics, local fields, etc.) are incompatible in some physical sense, or to recast the theory in such a form that it provides a practical language for the description of elementary particle dynamics." (Streater and Wightman 1964, p. 1).

The quest for internal consistency is often paired with the search for representation theorems that realize the axiomatically defined abstract structures through concrete mathematical structures. For quantum field theory this has been achieved by means of algebraic quantum theory (Haag 1992) where von Neumann's (1932) irreducibility postulate of conventional Hilbert space quantum mechanics was relaxed in favour of infinite-dimensional observable algebras. A classic example for another representation theory is the realization of abstract Boolean algebras by a field of sets. Another class of examples is due to fundamental representation theorems in the theory of measurement (Suppes et al. 1989). And Tversky (1977) developed a famous representation theorem for the representation of similarity in terms of features.

## 5.1 Linguistic example

Before we can conclude this section, we will illustrate the difference between phenomenological and foundational research program by an actual example from linguistics. In the last couple of years, within theoretical and philosophical linguistics, a lot of work was devoted to the comparison between Chomsky's (1995) minimalistic program and the earlier approach of principle and parameters (Chomsky 1981). Here are some claims of how to contrast the two programs (we roughly follow the outline in Hinzen 2000):

- What Chomsky calls "minimalism" is the search for general explanatory principles such as computational efficiency and related economy principles. It is one thing to find certain effects and an appropriate description of them (Principles and Parameters) and it is another thing to ask why this should be as it is ("minimalism").

- The aim of "minimalism" is to derive the earlier developed generalizations, principles, and explanations from independently needed more general principles. The task is to eliminate merely technical solutions.
- The attempt of "minimalism" is to rationalize the domain under discussion rather than to describe it (as in "principles and parameters").
- "Minimalism" tries to make sense of the different facts. Listing of facts is one thing, understanding them is another thing. Obviously, the Husserlian idea of 'phenomenology' refers to Chomsky's foundational program of minimalism.
- The evolving theory should be free of inner contradictions and conflicts. This relates to the interpretational issue discussed earlier. Even when the current theoretical debate is far from providing mathematical insights about the underlying structure and algebraic properties,[19] it should not be disregarded as a substantive research prospect (Sternefeld 2012).

What we can learn from these discussions is that foundational approaches (such as minimalism) and phenomenological approaches (such as "principles and parameters") are different modes of inquiry rather than different theories. In this sense, foundational approaches are not in competition with phenomenological ones. The choice for one of the two approaches is a free decision determined by our interests; there is no way to argue against the "wrong" approach. Both have their justification. However, in the history of science it is sometimes more useful to pursue one direction and not the other. It further should be added that a sound distinction between foundational and phenomenological programs should focus on the issue of representational theorems. The development of representational theorems is one of the main tasks of foundational research. It develops independently motivated abstract structures and proves how they can be realized by particular concrete structures. The latter can be applied in phenomenological research for describing particular phenomena and sensations.

### 5.2 A historical note about complementarity

One of the most important ideas of quantum mechanics is the concept of complementarity. Originally, the idea came from the psychology of consciousness, in particular the writings of William James:

> It must be admitted, therefore, that in certain persons, at least, the total possible consciousness may be split into parts which coexist but mutually ignore each other, and share the objects of knowledge between them. More remarkable still, they are complementary. (James 1890)

As noted by Max Jammer (1989), Nils Bohr, one of the founding fathers of quantum theory, was acquainted with the writings of James, and he has borrowed that idea from him.[20] In turn, Bohr introduced the idea into physics (complementarity of momentum and position), and he proposed to apply it beyond physics to human knowledge in general. However, his physical conception of complementarity is quite different from James', and his often-cited claim to apply it to human knowledge was never concretized by Bohr. In chapter VII of his book (James 1890) – on more than 10 pages – James describes several phenomena which illustrate the splitting of consciousness into parts that are not accessible from each other. For example, these phenomena concern the "unconsciousness in hysterics" (p. 202), partial blindness under "post-hypnotic suggestion" (p. 207) or the splitting of a person in several selves in "alcoholic delirium" (p.208). One example describes the common situation of partial anaesthesia:

> The mother who is asleep to every sound but the stirrings of her babe, evidently has the babe-portion of her auditory sensibility systematically awake. Relatively to that, the rest of her mind is in a state of systematized anaesthesia. That department, split off and disconnected from the sleeping part, can none the less wake the latter up in case of need. (p. 213)

---

[19] However, it should be noticed that beim Graben and Gerth (2012) presented a geometric representation theory for minimalist grammars in a mathematical sense.

[20] Similarly, in a letter to Stapp, Werner Heisenberg mentions "that Niels Bohr was very interested in the ideas of William James". (Stapp 1972, p. 1112)

Another example refers to the famous subject "Lucie" who was in a state of "post-hypnotic suggestion" and could see of all the cards covering her lap only those cards that were not a multiple of 3. She was particularly blind to numbers such as 9, 12, 15. Hence, the part consisting of the multiples of 3 was split off and disconnected from the part of numbers. However, under special conditions, when she had not to tell which cards she saw but to write it down by her hand, the other part of the numbers was accessible (p. 207).

Taking all the examples together, it seems adequate to use the term "autoepistemic accessibility" to refer to these phenomena. We use the term "autoepistemic" to refer to the epistemic states of a human subject who can reflect on her own epistemic states.[21] If two different states are not simultaneously epistemically accessible to the subject under discussion then they can be seen as complementary in James' sense.

It should be noted that the related term of "epistemic accessibility" has been introduced by beim Graben and Atmanspacher (2006) in a not so different sense. In their approach, the term refers to an external observer and his measurement apparatus for describing the "microstates" a dynamical system. These states are not accessible by macroscopic measurements which only provide a coarse-graining of the state space into "macrostates". In philosophical contexts, often the term *observer-dependency* is used for this phenomenon (Searle 1980, 1998). In this sense, different macroscopic measurement devices appear as complementary when they lead to such differences in epistemic accessibility. This phenomenon is pertinent in the neurosciences when different measurements such as fMRI or EEG generate different coarse-grainings of the brain's state space thus leading to incompatible descriptions of neurodynamics in the sense of Acacio de Barros and Suppes (2009). In the next subsection we will explain this idea in detail.

Bohr's concept of complementarity is clearly not copied from James' epistemic conception. As pointed out by Murdoch (Murdoch 1987, p. 55), the first occurrence of the word "complementarity" in Bohr's correspondence is in a letter to Pauli of August 13, 1927:

> What you write about your and Jordan's work on electrodynamics is extremely attractive and is very much in agreement with my own view about the nature of quantum theory, according to which the apparently contradictory requirements of superposition and individuality do not subsume contrary but complementary sides of nature. I am in complete agreement with your remarks on de Broglie's work: he is trying to achieve the impossible by a blending of two sides of the matter." (Cited from Murdoch 1987, p. 55).

---

[21] The term "autoepistemic" is widely used in (non-monotonic) logic, e.g. Moore (1988). Even when the reasoning component is not essential for the present treatment, one basic idea is present: an agent refers to his own epistemic states.

The terms "superposition" and "individuality" refer to the wave and particle, and in a draft version for a note to *Nature* referring to the wave-particle duality of light the word "complementarity" occurs in the very same sense:

> It seems that we here meet with an unavoidable dilemma, ... the question being not of a choice between two ~~different~~ rivalizing concepts but rather of the description of two complementary sides of the ~~same~~ phenomenon. (Cited from Murdoch 1987, p. 55. The emendations are Bohr's)

Bohr's conception of complementarity refers to the laws of nature rather than to the idea of (auto)epistemic accessibility as in James' writings. In other words, it is an ontic conception rather than an epistemic one.

Closely related to Bohr's concept of complementarity is Heisenberg's famous uncertainty principle.[22] In his book, *Die physikalischen Prinzipien der Quantentheorie*, Heisenberg (1944) starts his introduction of the uncertainty relation with another important idea of Nils Bohr: In order to find the right restrictions were the classical particle picture of physics can be applied one has to remember the idea that all facts of atomic physics that are describable in space and time have to be describable in the wave picture as well (p. 9). In the simplest case, a particle can be described in the wave picture by a 'wave packet'. However, for a wave packet no precise location and no precise velocity can be defined since the wave packet has a tendency to be dispersed over the whole space. According to the simple laws of optics the following uncertainty relation can be derived

(27) $\Delta q \cdot \Delta p \geq h$

Hereby, $\Delta q$ and $\Delta p$ denote the standard deviation (measuring the dispersion) of position and momentum, respectively. The constant $h$ is given as Planck's quantum of action relating radiation energy ($E$) to frequency ($f$) in the equation $E = h \cdot f$.

After explaining this optical picture of the uncertainty relation, Heisenberg makes clear that the uncertainty relations can be derived without reference to the wave picture by using the general schemes of quantum theory and its physical interpretation (Heisenberg 1944, p. 11). In fact, he provides a brief derivation of the principle (27) based on the matrix formalism of probability amplitudes. Generally, it is the complementarity of certain observables (expressed by their non-commutativity or order-dependence of the relevant operators) that allows the derivation of uncertainty relations. The general conclusion is that the uncertainty relations restrict the probability distributions predicted by the general formalism of quantum theory which apply to arbitrary physical states of the quantum system.

Yet another aspect of Heisenberg's discussion of the uncertainty relation deserves our attention. Obviously, irrespective of the concrete realization of physical states (as a wave packets or whatever), only the statistics of their measurements (or the probabilities of our expectations) satisfy this principle. In this sense, quantum theory can be seen as an abstract mathematical apparatus that acts like a kind of meta-theory and provides a new kind of language that allows to generalizing on a variety of particular mechanisms (partly known from optics and wave theory).

In Section 2 of his book, Heisenberg (1944) continues on this idea and provides several examples that demonstrate how measurements can change the state of a quantum system in order to satisfy the uncertainty relation. In this sense, the uncertainty relation formulates a property of the abstract meta-theory: there is no way to prepare a quantum state where the described object (electron, photon, etc.) has a sharp location and a sharp momentum at the same time.

---

[22] The term *uncertainty principle* is a translation of the German term *Unschärfeprinzip* or *Unbestimmtheitsprinzip*.

To say it in an embellished way: quantum theory is not really required and it is actually superfluous in a certain sense. You can get many of its results by cleverly combining classical mechanics, optics, and wave theory as had been carried out by the pioneers of naïve quantum physics, such as Einstein, Bohr, Sommerfeld, and de Broglie between 1905 and 1925. However, quantum mechanics provides a unified framework for formulating all these things more systematically. This hint can be helpful to understand the mathematical underpinnings of quantum theory for cognitive science. Again, the mathematical formalism is a kind of meta-theory that provides a useful language generalizing ideas known from neural network research, dynamic system theory, update semantics, and other frameworks.

Let us return now to the interpretational problem for probabilities. In quantum theory, a deep problem concerns the nature of the state vector. There are two basic questions the answers of which decide about its basic nature: (i) Does the state vector directly describe reality or is it related to our knowledge of the system?  (ii) Does the state vector describe a single object (particle or wave) or does is describe a whole ensemble of objects (or an ensemble of experiments with the object)? In modern quantum theory (contrasting with naïve quantum physics), the state vectors are connected with probabilities. However, there are three different interpretations of the whole concept of probability (e.g., Hájek 2012): frequency interpretation, subjective probability, and propensity interpretation. It is interesting to see how the different answers to the two basic questions concerning how state vectors relate to the involved concept of probability.

It was Max Born (1955) who gave a realistic answer to the first question, and he gave the ensemble answers to the second question. His concept of probability is clearly seeing probabilities as relative frequencies. There are deep conceptual problems with Born's approach, some of them are overcome by Ballentine's ensemble approach (Ballentine 1970).

Another realist conception assumes that the state vector describes a single object by giving probabilities a propensity interpretation (Popper 1959). Propensities can be assumed of being some kind of abstract, objective forces (unobservable dispositional properties) that provide a measure of the tendency of a situation to produce a certain event. Of course, Popper's article was an attack against Heisenberg's Copenhagen interpretation, which is based on a subjective interpretation of probabilities (reflecting our knowledge or our "consciousness" of particles).[23]

The third possibility is to assume an epistemic interpretation of the state vector and to assume that the state vector describes single objects. This clearly is the view of Heisenberg's Copenhagen interpretation with a subjective interpretation of probabilities. As made clear recently, this view does not necessarily entail observer-induced wave packet collapse (Barnum et al. 2000; Caves et al. 2002a, 2002b). More importantly, this picture conforms to the predominant picture of the Bayesian interpretation in Artificial Intelligence and Cognitive Psychology. Hence, it is suggestive to take this interpretation as the basic conception of probability in quantum cognition.

---

[23] Usually, Niels Bohr is taken as one of the main proponents of the "Copenhagen interpretation". However, as pointed out by Howard (2004), Bohr's view differs dramatically from Heisenberg's concerning the privileged role for the observer (positivism), in particular the observer-induced collapse of the state vector. Further, Heisenberg followed Einstein's leadership in reinterpreting ("umdeuten") basic concepts of physics, such as time, position and momentum. In contrast, Bohr followed Kant's philosophy in assuming that Human beings are endowed with the ability to think and imagine according to certain classical categories and schemas. We simply cannot get around the classical schemas in our thinking. Bohr concluded that in our thinking about the microscopic world we are forced to depend on classical categories and schemata grasping concepts such as position and momentum.

In the following section, we will concretize the notion of (auto)epistemic accessibility based on the operational interpretation of quantum physics. The operational setting suggests a particular algebraic structure for modelling propositions, one that is very different from the classical Boolean setting. The Boolean setting allows to model propositions as sets of possible worlds (with the operation of union, intersection and complement for the basic propositional operations. In contrast, the operational setting motivates a non-Boolean algebraic structure that invites to model propositions by subspaces of a Hilbert space or by projection operators of the Hilbert space and the corresponding lattice-theoretic operations. The algebra of propositions is defined by non-statistical axioms. Hence, the operational understanding does not require any notion of probability. The concept of probability will emerge by means of a measure function, its subjective interpretation can be motivated by a (quantum) de Finetti representation theorem (Barnum et al. 2000; Caves et al. 2002a, 2002b). Hereby, probabilities are taken to be degrees of belief, which are justified by axioms of fair bedding behaviour.

In this section, we have seen that James psychological conception of complementarity is clearly an epistemic one. In quantum physics, there is a big debate about the relationship between ontic and epistemic interpretations of complementarity, which will not concern us here further. Another historical lecture points to the abstract character of the mathematical formalism of quantum theory. It forms a kind of meta-theory providing a useful language for expressing certain generalizations, hopefully also in the domain of cognitive science.

# 6. Bounded rationality and the foundation of quantum probabilities

We come back now to the operational interpretation of quantum physics – an interpretation that is grounded in the reality of the process of measurement (contrasting with interpretations which assume real values for measurements before the measurements are actually carried out, before the questions are really asked). As we will see, this interpretation nicely fits with our concern of grounding the basic design features of quantum cognition. Instead of performing 'measurements' we confront our subjects with 'questions' in cognitive science. And in both cases we notice the 'answers' of our system and get a probabilistic outcome.

The mentioned design features can be based on a fairly abstract axiomatic framework, as found in the mathematical language of formal logic (as pioneered by Birkhoff and von Neumann (1936)). These features are free from any direct concern to a statistical or probabilistic interpretation. They concern what sometimes is called 'general quantum physics' (Piron, 1972) or 'generalized quantum theory' (Atmanspacher et al. 2002). This is a passage illustrating the crucial issue:

> "It is also clear that a satisfactory axiomatic structure of the kind referred to above cannot be formulated *a priori* in terms of wave functions, since their use would imply a statistical interpretation at the outset. The role of the wave functions in general quantum physics must emerge from the analysis of the more fundamental theory. As said above, the linear structure of the Hilbert space does appear, without reference to any statistical notions, as the appropriate description of general quantal systems (a set consisting of a family of Hilbert spaces describes systems which are not purely quantal, and the purely classical limit is described by a family of trivial one-dimensional Hilbert spaces). It is in this way that the statistical interpretation for wave mechanics will emerge as a consequence of essentially nonstatistical axioms, and what is presupposed in classical physics is clearly brought into evidence." (Piron 1972)

6.1 Operational realism

Classical (Kolmogorov) probability is based on a certain sample space *S* (also called the set of possible states/worlds). In this picture, propositions are considered to be subsets of *S*. The notion of a 'random variable' is of particular importance for classical theory. A random variable $\hat{\alpha}$ on a sample space *S* is a function from *S* onto some range $\alpha$ (called the test range); for example, $\alpha$ can be a set of arbitrary 'symbols' or a set of natural or real numbers. The proposition that the value of the random variable $\hat{\alpha}$ is *x* (for some $x \in \alpha$) can be defined as follows:

(28)    $\pi(\hat{\alpha} = x) =_{def} \{s \in S : \hat{\alpha}(s) = x\}$

It is easy to see that the system of propositions generated by $\pi(\hat{\alpha} = x)$ considering all $x \in \alpha$ defines a partition of the sample space *S*.

Similarly to the literature on test spaces (Wilce 2009), we consider all subsets *X* of a test range $\alpha$ and call these subsets 'events' of the random variable $\hat{\alpha}$. With each of these events a particular proposition (also called logical event) is connected; this is the proposition that the value of the random variable $\hat{\alpha}$ is an element of the considered subset *X*:

(29)    $\pi(\hat{\alpha} \in X) =_{def} \{s \in S : \hat{\alpha}(s) \in X\}$

If we consider the set of all propositions generated by $\pi(\hat{\alpha} \in X)$ for any $X \subseteq \alpha$ (including the empty set), and if we order these propositions by the subset relation, then we get a *Boolean lattice*. We call these propositions *experimental* or *testable* propositions. Yet this Boolean lattice is only a partial one, i.e. not all subsets of *S* are testable propositions.

The interesting question is what happens when we consider several of such partial Boolean lattices (defined by different random variables) and form the 'union' of these lattices. The obvious answer is that we do not get a Boolean lattice again. Instead, we obtain a so-called 'orthomodular' lattice. An orthomodular lattice is weaker than a Boolean lattice. Technically, it is a bounded lattice (the bounds are 0, 1) with join $\vee$ and meet $\wedge$, and on the lattice a unary operation ' (ortho-complementation) is defined such that, for all elements *x* and *y* of the lattice:

(30)    a.    *x''* = *x*
        b.    if *x* $\leq$ *y* then *y'* $\leq$ *x'*
        c.    *x* $\wedge$ *x'* = 0
        d.    if *x* $\leq$ *y* then *y* = *x* $\vee$ (*x'* $\wedge$ *y*) (orthomodular law)

Two elements *x* and *y* are called orthogonal iff *x* $\leq$ *y'*. An orthomodular lattice is a Boolean lattice, if in addition we have distributivity: *x* $\vee$ (*y* $\wedge$ *z*) = (*x* $\wedge$ *y*) $\vee$ (*x* $\wedge$ *z*). It can be shown (e.g. Piron 1972) that every orthomodular lattice is the union of its maximal Boolean sublattices (called *blocks*). Jenca (2001) provides some useful generalization of this standard finding. The following example – first introduced und discussed by Foulis and colleagues (Foulis and Randall 1972; Foulis 1999) – gives a handy illustration of the basic ideas. It defines the firefly box and its event logic.

Assume that there is a *firefly* erratically moving inside the box depicted in Figure 3. The box has two translucent (but not transparent) windows, one at the front and another one at the right. All other sides of the box are opaque. In principle, the firefly can be situated in one of the four quadrants {1,2,3,4}, and the firefly can be blinking or not (the latter is indicated by being in world 5).

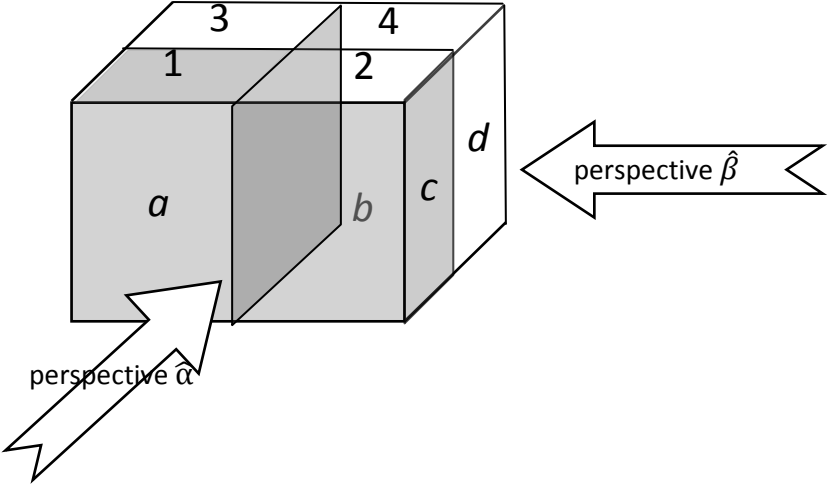

Figure 3. The firefly box. From perspective $\hat{\alpha}$, left view (worlds 1 and 3) corresponds to the half space *a* and right view (worlds 2 and 4) corresponds to the half space *b*. From perspective $\hat{\beta}$, however, left view (worlds 1 and 2) corresponds to the half space *c* and right view (worlds 3 and 4) corresponds to the half space *d*, by contrast.

For an external observer, the position of the firefly in one of the four quadrants is not visible even if the firefly is blinking. For testing whether the firefly is blinking and where it is the external observer can take one of two perspectives:

$\hat{\alpha}$: looking at the front window. If no blinking can be seen the outcome is *n*; if the blinking is at the left part the outcome is *a*; if the blinking is at the right part the outcome is *b*.

$\hat{\beta}$: looking at the side window: again no blinking is registered by *n*; if the blinking is at the left part the outcome is *c*; if the blinking is at the right part the outcome is *d*.

These two perspectives can be described by the sets $\alpha$ = {*a*, *b*, *n*} and $\beta$ = {*c*, *d*, *n*}. As already noted, all subsets of $\alpha$ generate a (partial) Boolean algebra of events. The same holds for all subsets of $\beta$. The results for testing {*a*, *b*} and {*c*, *d*} will be always the same – they are epistemically equivalent (beim Graben and Atmanspacher 2006, 2009). We can express this fact by saying that both events realize the same experimental proposition: $\pi${*a*, *b*} = $\pi${*c*, *d*}. Further, calling *X'* the complement of the event *X*, we can postulate that $\pi${*n'*} = $\pi${*a*, *b*}. Hence, for both perspectives we have eight experimental propositions: π{a}, π{b}, π{n}, π{a, b} = π{n'}, π{a, n} = π{b'}, π{b, n} = π{a'}, π{a, b, n} = **1** and π{} = **0**.

Two experimental propositions π{X} and π{Y} are called *compatible* iff the sub-lattice generated by {π{X}, π{Y}, π{X'}, π{Y'}} is distributive. For instance π{a} and π{b} are compatible but π{a}, π{c} are not. Experimental propositions that are not compatible are called *complementary*. Hence, π{a} and π{c} are complementary and so are π{a'} and π{c'}. As a matter of fact, two propositions of different blocks (i.e., generated by two different perspectives), where one does not contain the other informationally, are always complementary.

We get an explicit expression of the experimental (testable) propositions as sets of situations, and a set-theoretic idea of their identity and inclusion conditions, when we formulate two random variables $\hat{\alpha}(s)$ and $\hat{\beta}(s)$ as defined in Table 1. With this instrument we can see, for instance, that π{a} = {1,3}, π{a, b} = {1,2,3,4}, and π{c, d} = {1,2,3,4}. Hence, we obtain π{a, b} = π{c, d} and π{a} ≤ π{a, b} etc.

| Possible states $s$ | 1 | 2 | 3 | 4 | 5 |
|---|---|---|---|---|---|
| $\hat{\alpha}(s)$ | a | b | a | b | n |
| $\hat{\beta}(s)$ | c | c | d | d | n |

Table 1. Two perspectives to model the firefly box by using random variables.

Figure 4 shows the two arising Boolean lattices of testable propositions based on the two perspectives.

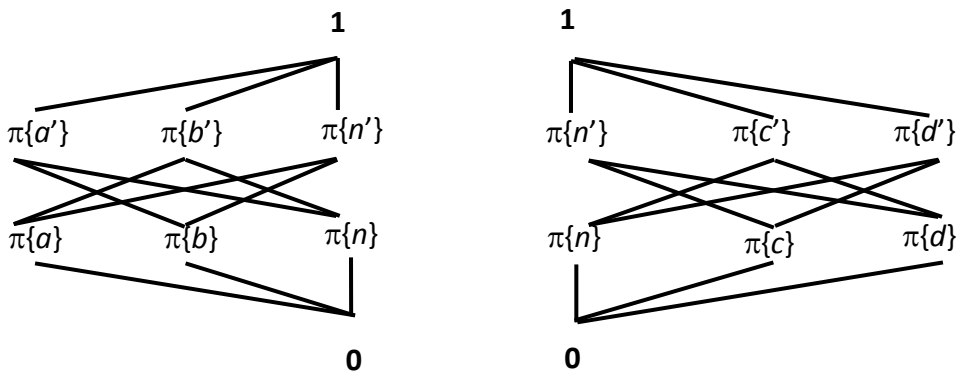

Figure 4. Two Boolean lattices resulting from the event algebras of perspective $\hat{\alpha}$ (left part) and perspective $\hat{\beta}$ (right part), respectively.

Now consider the union of the two perspectives which results in the lattice shown in Figure 5.

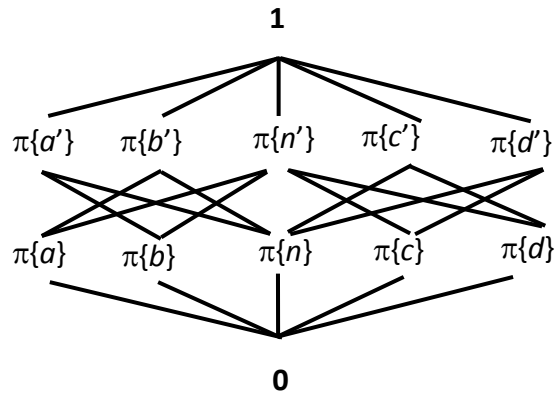

Figure 5. The union of the two Boolean lattices from Figure 4

The resulting lattice is orthomodular but not Boolean. The non-Boolean character is a consequence of the violation of distributivity. For instance, consider $\pi\{a\} \vee (\pi\{a'\} \wedge \pi\{d'\}) = \pi\{a\} \vee \pi\{n\} = \pi\{b'\}$. Distributivity would predict the equivalence with the term $(\pi\{a\} \vee \pi\{a'\}) \wedge (\pi\{a\} \vee \pi\{d'\})$. However, both parts of the conjunction give **1**, and hence the whole expression comes out as **1**. Since $\pi\{b'\}$ is different from **1**, distributivity is violated.

Atoms are events $A \neq 0$ such that there is no event $x \leq A$ unless $x = 0$ or $x = A$. A lattice is called *atomistic* iff for each element $x \neq 0$ of the lattice there exists some atom $A \leq x$. The *atomic covering law* states that for any atom $A$ the event $A \vee x$ covers $x$ (i.e. no element of the lattice lies strictly between $A \vee x$ and $x$). Wilce (2012) discusses several arguments that could help to motivate atomic covering. The example illustrated in Figure 4 is obviously atomistic and satisfying the atomic covering law. Yet there is another condition – irreducibility. An orthomodular lattice is called irreducible iff it cannot be expressed as a non-trivial direct product of simpler orthomodular lattices. Again, our firefly example lattice satisfies this condition.

Atomistic, irreducible orthomodular lattices satisfying the atomic covering law are called Piron lattices (after Piron, 1972 who was the first investigating such lattices). What we get by considering several random variables, defining the partial Boolean lattices of experimental propositions defined by each of these random variables, and considering the union of all these partial Boolean lattices, is a Piron lattice indeed. How do Piron lattices relate to Hilbert spaces? Interestingly, the lattice comprised by all sub-vector spaces of a given Hilbert space is equivalent to a Piron lattice. The atoms of this Piron lattice are the one-dimensional Hilbert spaces, of course.

The crucial question arising now is: how close is the connection between Piron lattices and projection lattices on Hilbert spaces? All projection lattices are Piron lattices. Unfortunately, the converse is not true: not every Piron lattice can be represented by a corresponding projection lattice. It needs additional conditions in order to prove the corresponding representation theorem (Solér 1995; Holland 1995). These conditions are rather technical and concern the infinite case. If they are satisfied, we call the lattice a Piron-Solér lattice. In case of the firefly box the corresponding lattice representation theorem entails the following three-dimensional vector space where the one-dimensional subspaces representing the atoms of the lattice are indicated by the symbols used in Figure 3:

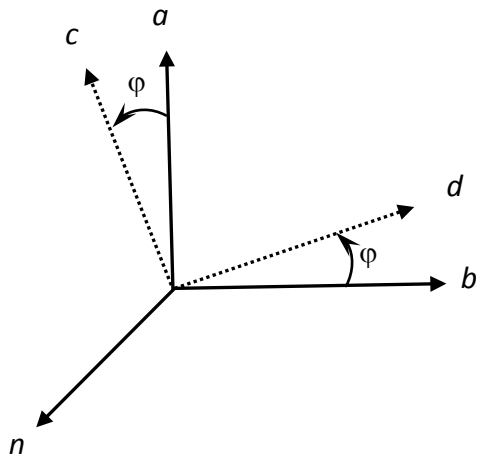

Figure 6. Hilbert space representation of the firefly box representing the two Boolean blocks $\alpha = \{a, b, n\}$ and $\beta = \{c, d, n\}$. The atoms $a, b, n$ (and $c, d, n$, respectively) are represented by pairwise orthogonal axes. The angle $\varphi$ between the complementary elements $a, c$ (and $b, d$, respectively) cannot be derived by the representation theorem from the orthomodular lattice and depends explicitly on a chosen probability model.

We see that in a Hilbert space representation compatible experimental propositions are represented by commuting projection operators while complementary experimental propositions are represented by operators that do not commute. Moreover, the complementation operator $X'$ is represented by Hilbert space orthocomplementation. Note further that the elements of the sample space $S$ have the ontological status of abstract objects. They help to express the identity conditions for experimental propositions. However, they are not necessarily represented as subspaces in the Hilbert space. For instance, worlds {1} and {2} do not correspond to some subspace, only their union {1, 2} does (corresponding to $c$). However, world {5} explicitly refers to the subspace denoted by $n$. In case that the overall structure of the Piron-lattice is Boolean, the sample space $S$ corresponds to an orthonormal basis of the Hilbert space. Hence, the classical case of a standard possible world semantics is a special case of the Hilbert space semantics.

In Section 4.1 we have introduced the notion of an additive measure function based on the lattice of projection operators of a given Hilbert space and we have shown that each unit state $u$ determines such a function of quantum probabilities on the testable propositions $A_i$ of the Hilbert space: $P_u(A_i) = ||A_i(u)||^2$. More generally, we have seen that also each mixture (or convex combination) of these probabilities is an additive measure function. Now the important question is whether each additive measure function can be represented by the mixture of quantum probabilities for pure states. The positive answer is the content of Gleason's theorem (Gleason 1957). Under the assumption that the dimension of the Hilbert space is larger than two, it states that each countably additive measure function can be expressed as the mixture (convex hull) of quantum probabilities for pure states (the latter following the Born rule, i.e. calculating the squared length of the projections of a given state).[24] For details, the reader is referred to the original paper by Gleason, and for a constructive proof see Richman & Bridges (1999).

Summarizing, we have considered measurements as providing partial Boolean lattices (also called blocks), which can be assumed to be formed by experimental propositions. If two experimental propositions of different blocks are not identical but overlap, then they are called *complementary*. Building the union of several blocks results in a structure called Piron-Solér lattice. The fundamental representation theorem states that Piron-Solér lattices and lattices of projectors of a Hilbert space are equivalent. The next step is to define additive measure functions on the space of experimental propositions – either defined as proposed in test theory or defined as subspaces of a Hilbert space, i.e. projection operators. Gleason's theorem tells us that probabilities arise either from the Born rule or their convex combinations. In this way, quantum probabilities are based on the lattice of experimental propositions. In physics, this lattice is constituted by the algebra of complementary observables. In the next subsections, we will consider how the related lattice can be motivated in cognitive science using a dynamic venue and this leads us back to the nature of bounded rationality.

## 6.2 Dynamical systems and symbolic dynamics

So far, the firefly box discussed in the previous subsection only provides a static picture about experimentally testable propositions. Observing the firefly's motion inside the box over an extended period of time would definitely deliver additional information about its actual position. The question is then, whether this additional information could be used to restore a classical Kolmogorovian description in the limit of infinite observation time. In order to deal with this question, we have to discuss some basic issues of dynamical system theory (see e.g. Ott 1993).

The firefly box can be regarded as a *deterministic dynamical system* with the firefly's actual position (namely three Cartesian coordinates $x = [x, y, z]$) inside the box as its *state* $x(t)$ at time $t$. The set of all possible states (i.e. of all possible spatial localizations) is called the *phase space X*.[25] Starting with some time $t_0$ the state $x(t_0)$ is called *initial condition*. The motion of the firefly is given by a one-parameter map $\Phi^t: X \to X$, the *phase flow* with parameter time $t \in [t_0, \infty)$, such that $x(t_0 + t) = \Phi^t(x(t_0))$. Following the states $x(t)$ for all times in the real interval $[t_0, \infty)$, yields the firefly's *trajectory*.

The unobserved firefly has a classical trajectory exploring the phase space *X*. As the firefly can only be observed when blinking, we get a temporal discretization of its continuous trajectory. In dynamical systems theory such discretization can be identified with stroboscopic or so-called Poincaré mappings. Finally, we only allow observational access either through the front window (perspective $\hat{\alpha}$), or through the side window (perspective $\hat{\beta}$), but not through both windows simultaneously. Again, these perspectives provide two complementary partitions of the phase space *X* into pairwise disjoint sets.

---

[24] As noted in Wilce (2012), with the help of quantum probabilities (based on Gleason's theorem) it is possible to reconstruct the rest of the formal apparatus of quantum mechanics, "the representation of 'observables' by self-adjoint operators, and the dynamics (unitary evolution). The former can be recovered with the help of the Spectral theorem and the latter with the aid of a deep theorem of E. Wigner on the projective representations of groups." (Wilce 2012)

[25] Of course, we could identify the points of the phase space with possible worlds. However, the idea of possible words is more general and takes an observer-dependent view (Searle 1998). In fact, the architect of a cognitive system can decide whatever he will consider as possible worlds, including entities that refer to the elements of a certain partition of the phase space (as we did in the previous section regarding the worlds 1, 2, 3, 4).

In order to illustrate the following argumentation let us crucially simplify the firefly box as follows: Assume the firefly's motion is confined to a two-dimensional phase space, the square [-1,1] × [-1,1]. Assume further that the firefly moves with constant speed along a closed circle of radius *R*=1, such that it reaches its initial state $x(t_0)$ after every period *T* again. Moreover, we suppose that the firefly also blinks periodically with some ratio $T_B = qT$, ($0 \leq q \leq 1$). For this two-dimensional firefly box, the front window coincides with the x-axis while the side window coincides with the y-axis. These views lead to two complementary partitions of the phase space *X* into pairwise disjoint sets *a* = [-1,0] × [-1,1], *b* = [0,1] × [-1,1] for perspective $\hat{\alpha}$, and *c* = [-1,1] × [-1,0], *d* = [-1,1] × [0,1] for perspective $\hat{\beta}$. Figure 7 displays the situation.

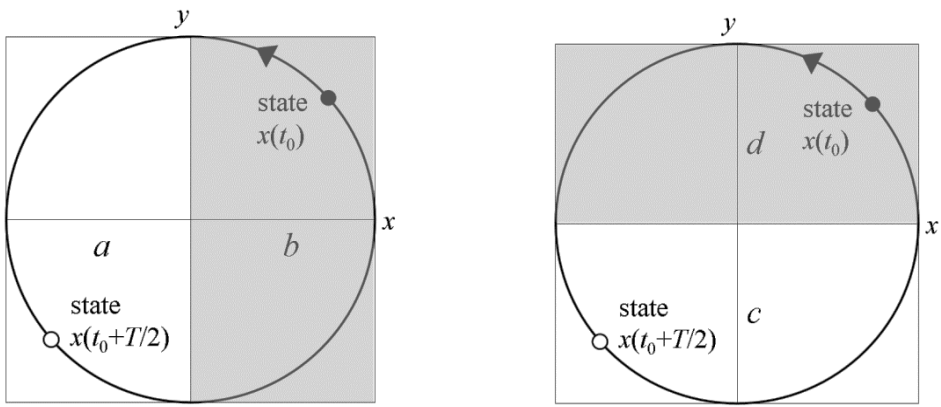

Figure 7. Phase space of the two-dimensional firefly box with circular trajectory. Partitions are α = {*a, b*} (left) and β = {*c, d*} (right). Time discretization ("blinking") with half period *T*/2.

So far, we have explained the idea of a (deterministic) dynamic system. Standard examples are classical mechanics describing the evolution of states of the phase space (pairs of location and momentum for one-particle systems). In neural network research, the *fast dynamics* of single neurons/neural networks can be seen as another dynamic system. It describes the evolution of neuronal activation for a single neuron or the activation spreading in neural networks.

Let us turn now to *symbolic dynamics*, which is a useful method for studying discrete-time dynamical systems with continuous state space. The crucial idea is to partition the state space into a finite number of subsets and to label each of these subsets with some symbol. In case of the firefly, the discretization of time through stroboscopic blinking and the discretization of phase space through partitioning, yields a symbolic representation of the system's trajectory as a sequence of partition cells being visited. If there is more than one observational perspective, more than one partition (and labelling function) will be involved.

In Figure 7 the initial condition belongs to rectangle *b* with respect to partition (perspective) α and to rectangle *d* with respect to partition β. Considering partition α first, we assign the symbol "*b*" to the initial condition $x(t_0)$. When the system evolves as indicated by the arrow in Figure 7, its state belongs to rectangle *a* after a half period *T*/2 when blinking appears again. Thus, we can assign the sequence "*ba*" to the first two iterations. When our firefly still blinks every half period, the complete trajectory is mapped onto an infinite string "*babababa*…" which is called a *symbolic dynamics* (Hao 1989; Lind and Marcus 1995). On the other hand, with respect to partition β, we assign the symbol "*d*" to the initial condition $x(t_0)$. The state at time $t_0$ + *T*/2 is contained in rectangle *c*, such that the corresponding string is "*dc*" for the first two iterations, eventually leading to the symbolic dynamics "*dcdcdcdc*…" for the given stroboscopic discretization.

Symbolic dynamics provides very powerful tools for discussing complex nonlinear dynamical systems when "unessential details" are captured by the coarse-graining of the phase space into partition cells. It allows investigating symbolic sequences by means of formal language and automata theory (Crutchfield 1994; Jiménez-Montaño et al. 2002; Moore 1990), information theory and Markov chains (Crutchfield and Feldman 2003; Crutchfield and Packard 1983) and algebraic quantum theory as well (Matsumoto 1997; Exel 2004).

The concatenation operation of symbolic dynamics has a straightforward interpretation in terms of phase space propositions. Generally, the set of phase space points generating a given symbolic sequence when taken as initial conditions can be interpreted as the proposition represented by the sequence. In our first example, the sequence "*ba*" corresponds to the proposition $\Pi("ba") = \{ x: x \in b$ and $\Phi^{\frac{T}{2}}(x) \in a\}$. Since $\Phi^{\frac{T}{2}}(x) \in a$ iff $x \in \Phi^{-\frac{T}{2}}(a)$, we get the following proposition representing the sequence "*ba*": $\Pi("ba") = b \cap \Phi^{-\frac{T}{2}}(a)$. Here, $\Phi^{-\frac{T}{2}}(a) = \left\{ x : \Phi^{\frac{T}{2}}(x) \in a \right\}$ denotes the pre-image of the cell *a*.

In general, a symbolic sequence $s_1 s_2 \ldots s_n$ refers to the following proposition:

(31) $\quad \Pi(s_1 s_2 \ldots s_n) = \bigcap_{k=0}^{n-1} \Phi^{-kT_B}(s_{k+1})$.

Therefore, any string $s_1 s_2 \ldots s_n$ of finite length *n* corresponds to an intersection of pre-images of partition cells under the phase flow. Because the intersection of two sets is in general smaller than each of the original sets (unless one of them is a subset of the other), longer strings correspond to smaller sets of initial conditions. If this is indeed the case, one speaks about the *dynamic refinement* of a partition.

Next, we have to distinguish two important cases. In the first one, the sets of possible initial conditions in phase space become smaller and smaller for increasing string lengths, eventually shrivelling into a singleton set which contains exactly one initial condition for an infinitely long symbolic sequence. In this case, which is common for chaotic dynamics where different trajectories are exponentially divergent, one speaks about a *generating partition* since the original partition generates the complete phase space in the limit of everlasting observation time. In other words, in the limit of sufficiently long symbolic sequences, we can identify each point of the phase space by such a sequence. There is no information lost by using the symbolic labelling method. For our two-dimensional firefly box, we can easily construct a generating partition when the sampling frequency is an irrational number, say $T_B = \sqrt{2}T$. In this case, the iterations of $\Phi^{kT_B}(x(t_0))$ for integer $k$ form a dense set around the circular trajectory. Hence the pre-images eventually converge to some initial condition $x(t_0)$ at the circle, as depicted in Figure 8.

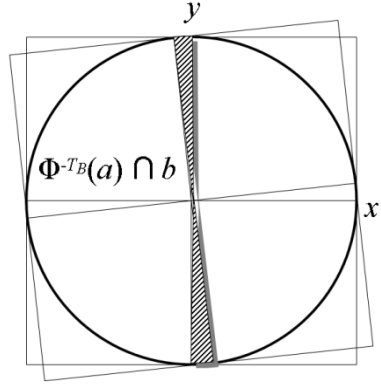

Figure 8. Generating partition of the two-dimensional firefly box with circular trajectory.

Thus, two generating partitions α and β both have the complete phase space *X* as their finest refinements such that their union is a Boolean lattice allowing for a Kolmogorovian probability theory.

In the second case, even the finest dynamic refinement of an initial partition exhibits residual coarse grains that cannot be further refined. These residuals correspond to the only propositions that are epistemically accessible with respect to a given perspective. Therefore, we obtain a Boolean partition algebra from the finest grains of an original partition. Beim Graben & Atmanspacher (2006, 2009) have argued that two different initial partitions α and β could yield two different Boolean partition algebras that are only partially overlapping. Then, our original firefly argumentation applies and the union of these algebras becomes an orthomodular lattice, leading to a canonical Hilbert space representation as required in quantum cognition.

The question whether any two partitions α, β are compatible or complementary to each other, depends crucially on the stroboscopic sampling. For a half period sampling $T_B=T/2$ we get the following pre-images: $\Phi^{-T/2}(a) = b$ and $\Phi^{-T/2}(b) = a$ one the one hand, and $\Phi^{-T/2}(c) = d$ and $\Phi^{-T/2}(d) = c$ one the other hand. As a consequence, the finest (symbolic) refinement of α is identically α, and the finest (symbolic) refinement of β is identically β (beim Graben et al. 2013). Therefore, both partitions are also dynamically complementary as no additional information is gained from the dynamic refinement.

By contrast, when we take $T_B=T/4$, the pre-images are $\Phi^{-T/4}(a) = d$ and $\Phi^{-T/4}(b) = c$ (cf. Figure 7). In this case, the sequence "*ba*", let say, is represented by the proposition Π("*ba*") = $b \cap \Phi^{-\frac{T}{4}}(a) = b \cap d$. This proposition corresponds to the first quadrant of Figure 7. Hence, the finest refinements of both α and β is the partition of the coordinate plane *X* into the four quadrants. For this discretization, the partitions α and β become compatible rather than complementary. One can express this situation by stipulating that the angle φ from Figure 6 will be an integer multiple of 90°.

Interestingly, our two-dimensional firefly box is completely equivalent to the phase portrait of a time-discretized harmonic oscillator where the spatial position is drawn along the x-axis while its velocity (actually its momentum) is drawn along the y-axis. Superimposing a cosine wave along *x* and a sine wave along *y* yields a circular trajectory as in Figure 7. Beim Graben et al. (2013) have intensively discussed the *epistemic quantization* (beim Graben & Atmanspacher 2006, 2009) of the harmonic oscillator. To this end, one has to consider probability distributions, so-called statistical states, of initial conditions in phase space. If these distributions are supported by the rectangles *a, b* or *c, d* from Figure 7, one can define statistical eigenstates either of position or of momentum by virtue of vanishing variance. Then, the same construction as for the two-dimensional firefly applies and a position eigenstate could not simultaneously be a momentum eigenstate and vice versa, which is exactly the situation of position/momentum complementarity in quantum physics.[26]

The firefly box and its dynamic counterpart, the harmonic oscillator, demonstrate that an epistemic quantization of an ontologically classical system is almost inevitable for coarse-grained descriptions. Such coarse-grainings are always due to limited precision and resolution of observational measurement devices. Sensory and perceptional apparatuses exhibit finite registration and relaxation times as well as finite resolution of their sampling ranges.

We illustrate this with our harmonic oscillator example again. Suppose that switching between alternating measurements of $\alpha$ and $\beta$ requires a quarter period *T*/4 measuring time. Observing the initial condition in rectangle *b* with respect to $\alpha$ as depicted in Figure 7, again, entails no information about its place in cells *c* and *d* with respect to $\beta$ whatsoever. Consequently, this uncertainty is propagated through a subsequent measurement of $\beta$ with respect to $\alpha$, since $\Phi^{T/4}(b) = d$.

In the previous section we found a Hilbert space representation of the firefly box exhibiting two Boolean blocks $\alpha$ = {*a, b, n*} and $\beta$ = {*c, d, n*}. In constructing this representation, a new parameter – the angle $\varphi$ between the complementary elements *a*, *c* (and *b*, *d*, respectively) – has been introduced. The question we will investigate now is whether symbolic dynamics can provide an interpretation of this parameter. As the first step in answering this question, we have to implement a probability model to the firefly box. Following Foulis (1999) again, this is achieved by a real-valued function $\omega$ that assigns numbers of the interval [0,1] to the propositions of the corresponding test range. Given an epistemically accessible statistical state, i.e. a probability distribution over the firefly box, the atomic propositions *a, b, c, d* gain the measures of their respective partition cells with respect to the original statistical state as probabilities. Thus, $\omega(S)$ with *S* = *a, b, c, d* provides a largely condensed static representation of the picture studied by beim Graben & Atmanspacher (2006, 2009). For perspective $\widehat{\alpha}$, the condition $\omega(a) + \omega(b) + \omega(n) = 1$ has to be satisfied on one hand, and for perspective $\widehat{\beta}$ the condition $\omega(c) + \omega(d) + \omega(n) = 1$, on the other hand. By means of continuation, this function becomes an additive probability measure over the proposition lattice depicted in Figure 5. The resulting space of *probability models* is convex, i.e., closed under convex combinations.

Therefore, the probability measure $\omega$ is uniquely determined by three real numbers

$$x = \omega(a); \quad y = \omega(c); \quad z = \omega(n)$$

as any other value, e.g. $\omega(b) = 1 - x - z$, can be obtained through convexity: $0 \leq x, y, z \leq 1$, $x + z \leq 1$, and $y + z \leq 1$.

---

[26] Moreover, beim Graben et al. (2013) coupled two harmonic oscillators together and showed that an eigenstate of a global observable could not simultaneously be an eigenstate of a local observable, a phenomenon known as *entanglement* in quantum physics. However, there is one crucial difference between complementarity and entanglement between coarse-grained classical dynamical systems and quantum systems: classical dynamical systems must be described by statistical states describing mixtures, i.e. states of ignorance; while proper quantum systems exhibit these phenomena already for pure states (which are individual phase space points for classical systems).

On the other hand, in our Hilbert space representation, Figure 6, the measure $\omega$ is given by a unit vector $w = [b, a, n]$ due to Gleason's theorem. Both descriptions can be mediated by the probability amplitudes $x = a^2, y = c^2, z = n^2$, such that the normalization constraint is identically fulfilled through convexity, via $b^2 = 1 - x - z$. However, for the complete characterization of the vector $w$ the rotation angle $\varphi$ in Figure 6 is required. This obeys the following constraint:[27]

$$\sqrt{x} \cos \varphi - \sqrt{1 - x - z} \sin \varphi = \sqrt{y}.$$

The solution is given as follows:[28]

(32) $\quad \varphi = \arccos \sqrt{\frac{y}{1-z}} - \arcsin \sqrt{\frac{1-x-z}{1-z}}$

The two-dimensional firefly moves continuously along the circle depicted in Figure 7 where it is not blinking most of the time. If we assume that blinking requires some amount of time, we can interpret $x$ as the proportion of blinking time in cell $a$ and $y$ as the proportion of blinking time in cell $c$. For the stroboscopic sampling with $T_B = T/2$ the firefly only blinks in cell $c$ when it is in cell $b$ as well. Therefore $\omega(b) = 1 - x - z = y = \omega(c)$ and hence $1 - z = x + y$. Inserting these numbers into equation (32) we obtain

(33) $\quad \varphi = \arccos \sqrt{\frac{y}{x+y}} - \arcsin \sqrt{\frac{y}{x+y}} = \frac{\pi}{2} - 2 \arcsin \sqrt{\frac{y}{x+y}}.$

In the special case of uniformly distributed blinking durations $x = y$ we get $\varphi = \frac{\pi}{2} - 2 \arcsin \sqrt{0.5} = \pi2 - 2\pi4 = 0$.

As a result, in the static firefly picture when dynamic refinement is not taken into account, the perspectives $\hat{\alpha}$ and $\hat{\beta}$ are compatible for uniform blinking time distributions. However, the general case could be realized when the firefly is attracted and repelled in cells $a$ and $b$ with different probabilities. In the dynamical systems framework this could be described by the presence of saddle nodes that are connected via heteroclinic sequences — an approach that has recently become increasingly popular in neural modelling (Rabinovich et al. 2008; beim Graben and Hutt 2014). Tuning the system's parameter, e.g. in such a way that $y = x(3 - 2\sqrt{2})$ yields $\varphi = \frac{\pi}{4}$, i.e. maximal incompatibility for the static firefly picture.

---

[27] Proof: Assuming a real Hilbert space $\mathcal{H}$, we identify the coordinates $b = \sqrt{1 - x - z}$ along the b-axis, $a = \sqrt{x}$ along the a-axis, and $n = \sqrt{z}$ along the n-axis in Figure 6. In the rotated coordinate system we have additionally $b = d \cos \varphi - c \sin \varphi$ such that $c = \sqrt{y}$ and, because $d$ is unknown, $d = b \cos \varphi + a \sin \varphi$. Eliminating $d$ from the first equation by means of the second one, yields after some rearrangements $b \sin \varphi = a \cos \varphi - c$, from which the conclusion follows.

[28] Proof: Using complex arithmetic for solving $a \cos \varphi - b \sin \varphi = c$ we obtain $\frac{a}{2}(e^{i\varphi} + e^{-i\varphi}) - \frac{b}{2i}(e^{i\varphi} - e^{-i\varphi}) = c$ and $\frac{1}{2}[ae^{i\varphi} + ae^{-i\varphi} + ibe^{i\varphi} - ibe^{-i\varphi}] = c$. Hence, $\frac{1}{2}[(a+ib)e^{i\varphi} + (a-ib)e^{-i\varphi}] = c$. Introducing the complex number $q = a + ib = re^{i\theta}$, with modulus $r = \sqrt{a^2 + b^2}$ and angle $\theta = \arctan(b/a)$, yields $\frac{1}{2}[re^{i\theta}e^{i\varphi} + re^{-i\theta}e^{-i\varphi}] = c$. Further, we get $\frac{r}{2}[e^{i(\theta+\varphi)} + e^{-i(\theta+\varphi)}] = c$ and $r \cos(\theta + \varphi) = c$. Hence, $\varphi = \arccos\left(\frac{c}{r}\right) - \theta$. Inserting $r$ and $\theta$ and utilizing another trigonometric identity entails the solution.

To summarize, four aspects make symbolic dynamics promising as a unifying framework of quantum cognition and neurodynamic systems theory. First, symbolic dynamics through coarse-graining normally leads to a loss of information. This is advantageous when the obtained abstraction captures essential traits of the original dynamics. There are many theoretical instruments to analyse symbol sequences, for example formal language theory, making symbolic dynamic especially powerful. Second, for many important systems *generating partitions* exist. In such a case, each point of the phase space can be uniquely identified by an infinite sequence of symbols. Studying the symbolic dynamics is therefore *completely equivalent* to studying the original dynamics. In other words, symbol-manipulating computations can be completely equivalent to continuous dynamics in such systems. In other cases yet where no generating partitions exist, we are confronted with complementary descriptions that still allow for a satisfactory symbolic description of the underlying continuous dynamics. Third, emergent symbolic dynamics can be stochastic for an underlying deterministic system. Important examples are Markov partitions for some chaotic deterministic systems that are generating. The resulting symbolic dynamics is a Markov process, which is equivalent to a stochastic system of random variables. Finally, fourth, symbolic dynamics can help to find interpretations of parameters that otherwise are left without any plausible interpretation. We have illustrated this in length with the hidden parameter $\varphi$ (introduced by Gleason's theorem) and its probabilistic interpretation above.

### 6.3 Bounded rationality and resource limitations

In Section 6.1, we have considered the general structure of propositions. We have argued that this structure conforms to orthomodular lattices. These lattices can be obtained through the union of (partial) Boolean lattices (or Boolean blocks) which conform to different perspectives, a cognitive agent can assume. According to James (1890), we regard these perspectives in correspondence to the different knowledge domains the agent is able to access. In this connection we introduced the term "autoepistemic accessibility". It is of primary interest to ask for the rationale behind the existence of different Boolean blocks. One possible hypothesis is that these blocks result from *cognitive resource limitations*. In the present section, we will check this hypothesis, and we will describe some arguments that illustrate how this idea could lead to empirical predictions.

First, we recognize an overall analogy between decision problems under bounded rationality and quantum cognition by comparing the prototypical decision situation depicted in Figure 1 with the firefly box from Figure 3. In Figure 1, a cognitive agent has to choose between two prospects *a* and *b*. Under prospect *a*, the two payoffs $x_1(a)$ and $x_2(a)$ are different from the corresponding payoffs $x_1(b)$ and $x_2(b)$, resulting from prospect *b*. In Figure 3, an observer can choose between two different perspectives $\hat{\alpha}$ and $\hat{\beta}$ of the firefly box. Taking perspective $\hat{\alpha}$, the possible outcomes are left view *a* and right view *b*, that are different from left view *c* and right view *d*, seen from perspective $\hat{\beta}$. Because the resources of the agent are limited with respect to observation timing and precision, she is unable to maintain both perspectives $\hat{\alpha}$ and $\hat{\beta}$ simultaneously – both perspectives lead to different partitions of the firefly box. Thus under bounded rationality, perspectives $\hat{\alpha}$ and $\hat{\beta}$ are complementary when their respective proposition lattices are Boolean blocks of a non-Boolean orthomodular (Piron-Solér) lattice comprised by their union. By means of this analogy, the prospects *a* and *b* of the original decision problem shown in Figure 1 could be complementary as well when the united propositional lattice turns out to be orthomodular. In this case, a Hilbert space representation could be constructed from which an instance of quantum cognition emerges. The crucial parameters of a quantum cognition model, phase angles and interference terms are entailed from a probability model by means of Gleason's theorem.

As a decent illustration of quantum cognition as bounded rationality, we discuss the phenomenon of speed-accuracy tradeoff (SAT) as described by Wickelgren (1977). The example we will consider here is the paradigmatic letter recognition task of Reed (1973). In Reed's classical experiment, subjects were presented a list of three consonants, e.g. 'bkr', they had to read aloud. Then a mask of a three-digit number appeared for some time that the subjects had to count down by three for 15 seconds. Afterwards, a probe consonant, say, 'k', appeared for some period of time that the subjects had to retrieve from the priming list. Plotting retrieval accuracy (discriminability *d'*) against total recognition time, exhibits characteristic, so-called SAT curves.

Up to now, decision times and SAT curves have been very successfully modelled by random walk or, in the continuum limit, drift-diffusion models (Busemeyer and Townsend 1993; Ratcliff 1978; Ratcliff and Rouder 1998). In order to describe Reed's findings in a random walk model, one introduces a variable for the accumulated preference of a binary decision. Say, 'k' was the probe that appeared in the last trial of the experiment, the subject has to decide whether 'k' had appeared in the list 'bkr' (YES) or whether it had not (NO). Because the correct response is YES, the amount of accumulated preference stochastically drifts towards a certain threshold until hitting either the YES or the NO threshold, signalling either a correct or an erroneous decision.

We fit Reed's (1973) original data to Ratcliff's (1978) drift-diffusion model in Fig. 9, where discriminability *d'* depends on time $t$ though

$$d' = \frac{d'_{\text{ASY}}}{\sqrt{1 + s^2/[\eta^2 \ (t - t_0)]}}$$

here $d'_{\text{ASY}}$ is the asymptotic discriminability, $s^2$ is drift variance, $\eta^2$ is relatedness variance, and $t_0$ is the origin of the timescale (i.e. intercept).

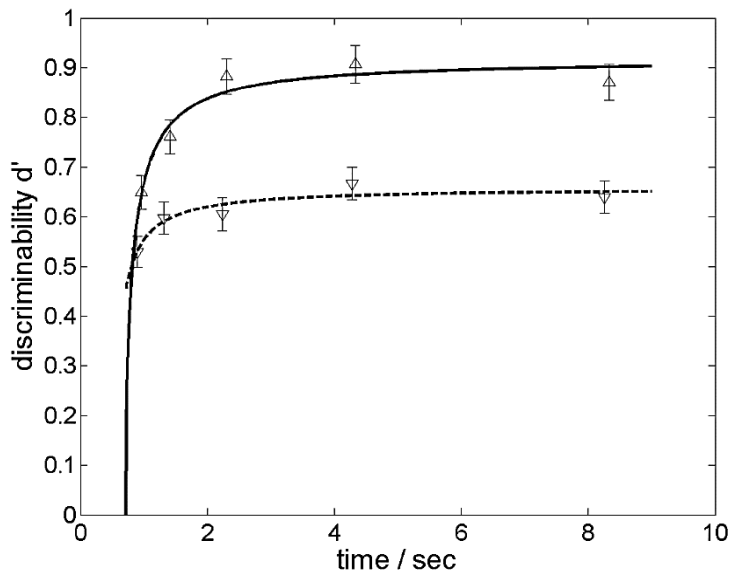

Figure 9. Speed-accuracy tradeoff (SAT) functions for the letter recognition experiment of Reed (1973). Discriminability *d'* estimated as complementary equivalent error rate (1-*r*) [Reed (1973), Fig. 2]. Two single subjects were investigated (called E.R. and N.V., respectively). Upper triangles and solid fit for subject E.R.; lower triangles and dashed fit for subject N.V. Error bars indicate one standard error.

The SAT functions in Figure 9 indicate that retrieval accuracy is the larger the more decision time the subject has available. In other words, the shorter the decision time the larger the retrieval error, which explains the notion speed-accuracy tradeoff.

Regarding theoretical understanding, Busemeyer and Townsend's (1993) *decision field theory*, addresses several pertinent problems of classical decision theory related to the simple model from Figure 1 and to the assessment of subjective expected utility – see equations (1a,b). One important step in this model introduces a random walk dynamics where the agent's preference towards prospect *a* against prospect *b* becomes accumulated over some period of time. In this sense, we can interpret the temporal accumulation of evidence in decision field theory by means of *virtual decisions* between complementary perspectives in terms of the quantum cognition framework presented in Section 6.2. In other words, even before the final decision takes place, the cognitive agent has to assess the subjective expected utility of competing prospects from an autoepistemic point of view. Under bounded rationality constraints this cannot be achieved simultaneously but must be carried out successively, when the agent's cognitive resources are too much limited. In this case, one prospect is autoepistemically not accessible from the chosen perspective of the other prospect, and vice versa. Hence the decision problem must be solved by accumulating preference through virtual decisions between complementary perspectives.

Traditionally, the SAT effect is induced by differences in the preparation of the experiment: A subject could be instructed either to be as fast or to be as accurate as possible (Wickelgren 1977). In the first case, time constraints can be prescribed as in Reed's (1973) experiment. In the second case, accuracy can be controlled by different payoffs. In a random walk model, both kinds of manipulation are eventually reflected by the positioning of the first-passage threshold, θ, triggering the decision. If θ is large, accuracy is emphasized on the expense of long first-passage times. If, on the other hand, θ is small, first-passage times become small as well on the expense of a large error rate. Figure 10 illustrates these two possibilities along the lines of Ratcliff & Rouder (1998) and Busemeyer and Townsend (1993). Here we simulate a random walk process (Busemeyer and Townsend 1993, Eq. (3b))

$$\hat{A}(t) = \hat{A}(t-1) + d + \hat{\varepsilon}(t)$$

where $\hat{A}(t)$ is the accumulated preference after sampling time $t$, $d$ is the drift whose sign indicates the direction of the decision dynamics: $d > 0$ for YES and $d < 0$ for NO, and $\hat{\varepsilon}(t)$, the residual, is a Gaussian random process with zero mean and variance $\sigma^2$.

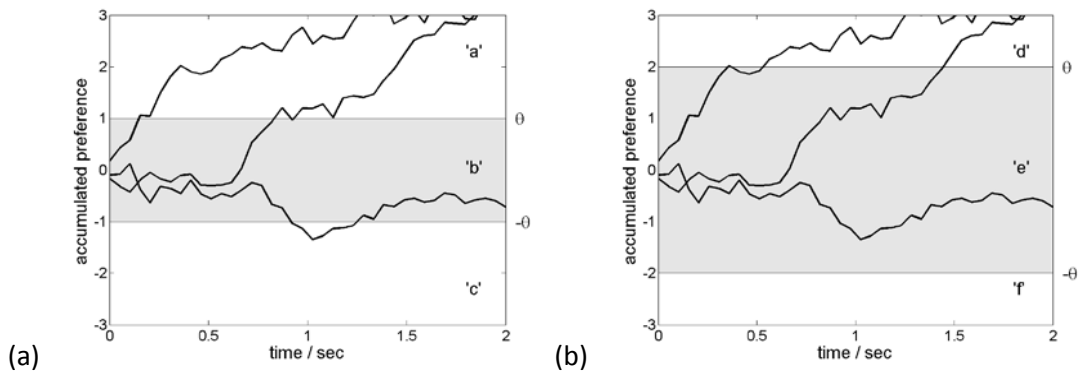

(a) (b)

Figure 10. Three realizations of a random walk model (Busemeyer and Townsend 1993) with positive drift $d$ = 0.05 and residual variance $\sigma^2$ = 0.2. (a) Small decision threshold |θ| = 1 (dashed lines) yields small first-passage times but error rate = 0.33. (b) Larger threshold |θ| = 2 (dashed lines) yields longer decision times, yet perfect accuracy of 1.0.

Looking at Figure 10 immediately suggests an interpretation of the SAT effect in terms of dynamical systems and symbolic dynamics: Accumulated preference $\hat{A}(t)$ is a scalar quantity comprising a one-dimensional phase space (in contrast to the two-dimensional example of the harmonic oscillator discussed above). Therefore, the y-axis in Fig. 10 can be interpreted as the system's phase space alone. A partition of the system's phase space is then any decomposition of the y-axis into disjoint intervals. Thus, the decision threshold θ partitions the scale of accumulated preference in three intervals. Assigning a symbol "*a*" to the interval [θ, ∞), "b" to the interval (-θ, θ) and "c" to the interval [-∞, -θ) of Fig. 10(a) creates a trinary symbolic dynamics (beim Graben & Kurths 2003) of the random walk model by virtue of the encoding rules $s_t$ = "*a*" for all times *t* larger than the first-passage time with respect to θ; $s_t$ = "c" for all times *t* larger than the first-passage time with respect to -θ; and $s_t$ = "b" for all other times *t*. Likewise, the different encoding threshold θ of Fig. 10(b), partitions the y-axis into cells symbol "d*"* for the interval [θ, ∞), "e" for the interval (-θ, θ) and "f" for the interval [-∞, -θ) with encoding rules $s_t$ = "d" for all times *t* larger than the first-passage time with respect to θ; $s_t$ = "e" for all times *t* larger than the first-passage time with respect to -θ; and $s_t$ = "f" for all other times *t*. The symbolic sequences resulting from these both encoding procedures are *eventually periodic*, thereby indicating bistable attractor dynamics. Yet it has been proven that partitions for multistable dynamical systems can never be generating (beim Graben 2004). Therefore, two different partitions that are induced by two different decision thresholds θ such as in Figs. 10(a) and (b), respectively, are incompatible to each other in the sense of the epistemic quantization of a classical dynamical system (beim Graben & Atmanspacher 2006, 2009). Hence we conclude that the SAT effect can be interpreted in terms of that (auto)epistemic quantization being compatible with its classical description by random walk processes.[29]

Summarizing we see that the coarse-grained random walk model of the SAT phenomenon leads naturally to an autoepistemic quantization of the agent's one-dimensional preference dynamics. Different experimental instructions of the subject yield different partitions of the subject's autoepistemic states which appear to be complementary since none of them is generating. Interpreting the sequences of the resulting symbolic dynamics as propositions, their lattices become Boolean blocks of a united orthomodular lattice. This, we conclude, is the foundation of quantum cognition from bounded rationality.

# 7. Quantum cognition and its formal grounding: some tentative conclusions

In the first part of this article, we have considered several puzzles of bounded rationality, and we have shown how the present account of quantum cognition – taking quantum probabilities rather than classical probabilities – can give a more systematic clarification of these puzzles than the alternate and rather eclectic treatments in the traditional framework of bounded rationality. Unfortunately, the quantum probabilistic treatment does not always and does not automatically provide a deeper understanding and a true explanation of these puzzles. The reason is that the quantum idea introduces several new parameters which possibly can be *fitted* to empirical data but which do not necessarily *explain* them. Hence, the phenomenological research has to be augmented by responding to deeper foundational issues. The second part of the paper is devoted to that problem.

The foundational approach to quantum cognition exploits ideas of the operational approach to quantum physics which does not take the Hilbert space as a given conceptual framework but rather which tries to motivate it. Following the research by Foulis, Randall and colleagues such a foundational framework is motivated by assuming partial Boolean algebras that describe the particular perspectives of a cognitive agent. These perspectives are combined into a uniform system while considering certain capacity restrictions. Technically, this gives an orthomodular-lattice – a structure that can violate distributivity. From an empirical perspective, it is at this point that one important aspect of the whole idea of bounded rationality directly enters the theoretical scenery of quantum cognition: resource limitation. Resource limitation has the effect that not all possible perspectives can be simultaneously maintained by a cognitive agent.[30]

---

[29] This interpretation does not exclude another possibility that is currently being investigated by several researchers, namely the need of quantum dynamics instead of classical Markov models for the description of the time-course of decision making processes (Busemeyer, Wang, & Townsend 2006, Busemeyer et al. 2009, Busemeyer and Bruza 2012). It might actually be tempting to speculate about the need of a *second quantization* of decision field theory to this end. However, at the present stage it is too early to make any definite statements about these recent developments.

[30] An interesting question that will arise is the problem of skill acquisition or automatization. The present view is compatible with an instance theory of automatization (Logan 1988) in which automatization is construed by the acquisition of separate representational instances.

One important issue both in physics and in the new field of quantum cognition concerns the notion of compatibility and complementarity. In physics, the complementarity of two observables has nothing to do with the incompetence or inability of the observer being able to perform two simultaneous measurements. Rather, it has to do with the nature of the measurements per se. We have seen already that in the cognitive realm observables correspond to questions that are asked to a human subject whose behaviour is investigated. Complementarity can be the result of the inability of our human subject to perform simultaneous tasks. In Section 6.3 we have illustrated how this idea helps to understand phenomena such as speed-accuracy trade-offs.

In discussing the Ellsberg puzzle another important aspect of the foundational issue came to light. The foundational argument is connected with Gleason's theorem and automatically leads to a distinction between probabilities that are defined by pure states and probabilities arising from the statistical mixture of such states. It is possible to relate this formal distinction to the deep conceptual distinction between risk and ignorance.

Curiously enough, there are proponents of cognitive science who set up an opposition between computation, conceived in the usual symbol-manipulation sense, and continuous nonlinear dynamics (e.g., Fodor and Pylyshyn 1988). This seems to us quite misleading, if only because the existence of generating partitions shows how symbol-manipulating computation can be *completely equivalent* to a dynamical system. In cases where we do not find generating partitions it still can be possible that complementary descriptions exist that give a satisfying symbolic description of the underlying continuous nonlinear dynamics. In this sense, the instruments of quantum description help to minimize the gap between symbolic descriptions and descriptions in terms of nonlinear differential equations as used in neuroscience.

According to beim Graben (2004), the transient dynamics of a nonlinear system in general and of a neural network in particular can be interpreted as *computation*, i.e. the manipulation of discrete symbols, upon a partition of the system's phase space. The resulting symbolic dynamics belongs to a certain complexity class and can be generated by a suitable dynamic automaton (beim Graben 2004). So far, partitions have been introduced as voluntary perspectives of an observer. However, they can also be created internally, from an autoepistemic point of view, which has been shown in the framework of *liquid computation* (Maass et al. 2002; beim Graben et al. 2009). A liquid computer is a large disordered neural network with random connectivity. Only a small sub-network of so-called *read-out neurons* is trained to solve prescribed classification tasks on the high-dimensional phase space trajectories, i.e. the read-out neurons partition the remaining phase space. If the read-out neurons are trained with more than one classification tasks, the resulting partitions may become complementary, hence yielding quantum cognition capabilities of neural networks (Acacio de Barros and Suppes 2009).

In quantum physics, we have an impressive series of *crucial experiments* that provide *prima facie* evidence for the basic assumption of quantum theory and the failing of classical theories. Prominent examples include the photoelectric effect (demonstrating particle properties of light), the Compton effect, the Frank-Hertz experiment, double slit experiments with single photons or with single electrons and Stern-Gerlach experiment. Textbooks of quantum physics give many other examples that provide a direct demonstration of quantum effects. In all these cases, the things speak for themselves and there is no need to provide a complicated reasoning process or to provide extraneous details, since any reasonable person would immediately find the facts convincingly demonstrating the quantum hypothesis.

Unfortunately, the situation in quantum cognition is much less convincing concerning *prima facie* evidence that proves the need of basic assumptions of quantum cognition. Critical experiments are rare in the field. Perhaps most convincing is what we have considered in Section 4.3 in connection with order phenomena, namely Wang's and Busemeyer's (2013) demonstration that the quantum approach can describe all four types of order effects that were considered in Section 3.4. Of special interest is their verification of the 'law of reciprocity'. In this case, each reasonable person will admit that classical probability cannot account for this constraint. Only the formalism of quantum theory is able to do so.

In the field of cognitive linguistic, the situation is much less clear. Though phenomena such as prototypicality, vagueness, polysemy, and invited inferences (Grice 1989) are cases in point to apply the formalism of quantum theory (e.g., Blutner 2009), there are only few phenomena that provide a rather direct evidence in favour of such a formalism. A potential example is the phenomenon of borderline vagueness (Blutner et al. 2013).

Summarizing, we conclude that the future of quantum cognition depends on the discovery of phenomena that provide *prima facie* evidence for crucial assumptions of quantum cognition. Perhaps, cognitive psychology and cognitive linguistics are potential areas for corresponding discoveries. However, also fields such as music theory – especially mathematical theories of tonal music including theories of modulation and emotional meaning could give new impulses (Mazzola 2002; Tymoczko 2011; Lerdahl 2001). For instance, structural quantum probabilities – arising from the structure of the Hilbert spaces – could be used to model universal traits of cognition that cannot be learned but have to be assumed as innate and emerging from an underlying Hebbian neurodynamics (Large 2010). The methodological aspects of founding quantum probabilities we have discussed in this paper could be a helpful guide on this road.

**Acknowledgement**

We would like to thank Harald Atmanspacher, Rens Bod, Jerome Busemeyer, Thomas Filk, Christopher Fuchs, Emmanuel Pothos, Remko Scha, Rüdiger Schack, and Sonja Smets for enlightening hints and useful conversation. One of us (RB) thanks Prof. Vishal Shani and Prof. Preem Saran Satsangi for a stimulating and superb stay at the Dayalbagh Educational Institute (Agra, India). PbG greatfully acknowledges support from the German Research Foundation DFG through a Heisenberg grant (GR 3711/1-2).

**References**

Acacio de Barros, J., & Suppes, P. (2009). Quantum mechanics, interference, and the brain. *Journal of Mathematical Psychology, 53*(5), 306-313.

Aerts, D. (1982). Example of a macroscopical classical situation that violates Bell inequalities. *Lettere Al Nuovo Cimento, 34*(4), 107-111.

Aerts, D. (2009). Quantum structure in cognition. *Journal of Mathematical Psychology, 53*, 314-348.

Aerts, D., Czachor, M., & D'Hooghe, B. (2005). Do We Think and Communicate in Quantum Ways? On the Presence of Quantum Structures in Language. In N. Gontier, J. P. V. Bendegem, & D. Aerts (Eds.), *Evolutionary Epistemology, Language and Culture.* (Studies in Language, Companion series). Amsterdam: John Benjamins Publishing Company.

Aerts, D., & Gabora, L. (2005). A state-context-property model of concepts and their combinations II: A Hilbert space representation. *Kybernetes, 34*(2), 176-205.

Aerts, D., & Sassoli de Bianchi, M. (2014a). The unreasonable success of quantum probability I: Quantum measurements as uniform fluctuations. *arXiv preprint arXiv:1401.2647*.

Aerts, D., & Sassoli de Bianchi, M. (2014b). The unreasonable success of quantum probability II: Quantum measurements as universal measurements. *arXiv preprint arXiv:1401.2650*.

Allais, M. (1953). Le comportement de l'homme rationnel devant le risque: Critique des postulats et axiomes de l'ecole Americaine. *Econometrica: Journal of the Econometric Society, 21*, 503-546.

Atmanspacher, H., & Römer, H. (2012). Order effects in sequential measurements of non-commuting psychological observables. *Journal of Mathematical Psychology, 56*, 274-280.

Atmanspacher, H., Römer, H., & Walach, H. (2002). Weak Quantum Theory: Complementarity and Entanglement in Physics and Beyond. *Foundations of Physics, 32*(3), 379-406.

Axler, S. (1996). *Linear Algebra done right*. New York, Inc.: Springer-Verlag.

Ballentine, L. E. (1970). The statistical interpretation of quantum mechanics. *Reviews of Modern Physics, 42*(4), 358-381.

Baltag, A., & Smets, S. (2011). Quantum logic as a dynamic logic. *Synthese, 179*(2), 285-306.

Barnum, H., Caves, C. M., Finkelstein, J., Fuchs, C. A., & Schack, R. (2000). Quantum probability from decision theory? *Proceedings of the Royal Society of London. Series A: Mathematical, Physical and Engineering Sciences, 456*(1997), 1175-1182.

beim Graben, P. (2004). Incompatible Implementations of Physical Symbol Systems. *Mind and Matter, 2*(2), 29–51.

beim Graben, P. (2014). Order effects in dynamic semantics. *Topics in Cognitive Science, 6*(1), 67-73.

beim Graben, P., & Atmanspacher, H. (2006). Complementarity in classical dynamical systems. *Foundations of Physics, 36*(2), 291-306.

beim Graben, P., & Atmanspacher, H. (2009). Extending the philosophical significance of the idea of complementarity. In H. Atmanspacher, & H. Primas (Eds.), *Recasting Reality. Wolfgang Pauli's Philosophical Ideas and Contemporay Science* (pp. 99-113). Berlin: Springer.

beim Graben, P., Filk, T., & Atmanspacher, H. (2013). Epistemic entanglement due to non-generating partitions of classical dynamical systems. *International Journal of Theoretical Physics, 52*, 723-734.

beim Graben, P., & Gehrt, S. (2012). Geometric representations for minimalist grammars. *Journal of Logic, Language and Information, 21*, 393-432.

beim Graben, P., & Hutt, A. (2014). Attractor and saddle node dynamics in heterogeneous neural fields. *EPJ Nonlinear Biomedical Physics, 2*(1), 4.

Birkhoff, G., & von Neumann, J. (1936). The logic of quantum mechanics. *Annals of Mathematics, 37*(4), 823-843.

Blutner, R. (2009). Concepts and Bounded Rationality: An Application of Niestegge's Approach to Conditional Quantum Probabilities. In L. Accardi, G. Adenier, C. Fuchs, G. Jaeger, A. Y. Khrennikov, J.-Å. Larsson, et al. (Eds.), *Foundations of Probability and Physics-5* (Vol. 1101, pp. 302-310). New-York: American Institute of Physics Conference Proceedings.

Blutner, R. (2012). Questions and Answers in an Orthoalgebraic Approach. *Journal of Logic, Language and Information, 21*(3), 237-277, doi:10.1007/s10849-012-9158-0.

Blutner, R., Hendriks, P., & de Hoop, H. (2003). A new hypothesis on compositionality. In P. P. Slezak (Ed.), *Proceedings of the Joint International Conference on Cognitive Science* (pp. 53-57). Sydney: ICCS/ASCS.

Blutner, R., Hendriks, P., De Hoop, H., & Schwartz, O. (2004). When Compositionality Fails to Predict Systematicity. In S. D. Levy, & R. Gayler (Eds.), *Compositional Connectionism in Cognitive Science. Papers from the AAAI Fall Symposium* (ISBN 1-57735-214-9 ed., pp. 6-11). Arlington: The AAAI Press.

Blutner, R., & Hochnadel, E. (2010). Two Qubits for C.G. Jung's Theory of Personality. *Cognitive Systems Research, 11*(3), 243-259, doi:10.1016/j.cogsys.2009.12.002.

Blutner, R., Pothos, E. M., & Bruza, P. (2013). A Quantum Probability Perspective on Borderline Vagueness. *Topics in Cognitive Sciences, 5*, 711-736.

Born, M. (1955). The statistical interpretation of quantum mechanics. *Science, 122*.

Bruza, P., Busemeyer, J. R., & Gabora, L. (2009). Introduction to the special issue on quantum cognition. *Journal of Mathematical Psychology*.

Budescu, D., & Wallsten, T. (1995). Processing Linguistic Probabilities: General principles and Empirical Evidence. In J. R. Busemeyer, R. Hastie, & D. L. Medin (Eds.), *Decision making from a cognitve perspective (Vol. Psychology of Learning and Motivation, Volume 32)* (pp. 275-318). San Diego, CA: Academic Press.

Busemeyer, J. R., & Bruza, P. D. (2012). *Quantum Cognition and Decision*. Cambridge, UK Cambridge University Press.

Busemeyer, J. R., Pothos, E. M., Franco, R., & Trueblood, J. S. (2011). A quantum theoretical explanation for probability judgment errors. *Psychological Review, 118*(2), 193-218.

Busemeyer, J. R., & Townsend, J. T. (1993). Decision field theory: A dynamic-cognitive approach to decision making in an uncertain environment. *Psychological Review, 100*, 432 - 459.

Busemeyer, J. R., Wang, Z., & Townsend, J. T. (2006). Quantum dynamics of human decision-making. *Journal of Mathematical Psychology, 50*(3), 220-241.

Carbó, R. (Ed.). (1995). *Molecular Similarity and Reactivity: From Quantum Chemical to Phenomenological Approaches*. Dordrecht: Kluwer.

Caves, C. M., Fuchs, C. A., & Schack, R. (2002a). Quantum probabilities as Bayesian probabilities. *Physical review A, 65*(2).

Caves, C. M., Fuchs, C. A., & Schack, R. (2002b). Unknown quantum states: the quantum de Finetti representation. *Journal of Mathematical Physics, 43*, 4537 - 4559.

Chomsky, N. (1981). *Lectures on government and binding*. Dordrecht: Foris.

Chomsky, N. (1995). *The minimalist program*. Cambridge MA: MIT Press.

Chomsky, N. (2005). Three factors in language design. *Linguistic Inquiry, 36*(1), 1–22.

Conte, E. (1983). *Exploration of Biological Function by Quantum Mechanics.* Paper presented at the Proceedings 10th International Congress on Cybernetics,

Conte, E., Khrennikov, A. Y., Todarello, O., Robertis, R. D., Federici, A., & Zbilut, J. P. (2008). A Preliminary Experimental Verification on the Possibility of Bell Inequality Violation in Mental States. *NeuroQuantology, 6*(3), 214-221.

Crutchfield, J. P. (1994). The calculi of emergence: Computation, dynamics and induction. *Physica D, 75*, 11-54.

Crutchfield, J. P., & Feldman, D. P. (2003). Regularities unseen, randomness observed: Levels of entropy convergence. *Chaos and Complexity Letters, 13*, 25-54.

Crutchfield, J. P., & Packard, N. H. (1983). Symbolic dynamics of noisy chaos. *Physica D, 7*, 201-223.

de Hoop, H., Hendriks, P., & Blutner, R. (2007). On compositionality and bidirectional optimization. *Journal of Cognitive Science, 8*(2).

Ellsberg, D. (1961). Risk, Ambiguity, and the Savage Axioms. *Quarterly Journal of Economics, 75*(4), 643-669.

Engesser, K., Gabbay, D. M., & Lehmann, D. (Eds.). (2009). *Handbook of Quantum Logic and Quantum Structures*. Amsterdam: Elsevier.

Exel, R. (2004). KMS states for generalized gauge actions on Cuntz-Krieger algebras (An application of the Ruelle-Perron-Frobenius theorem). *Bulletin of the Brazilian Mathematical Society, 35*, 1-12.

Fodor, J. A., & Pylyshyn, Z. W. (1988). Connectionism and cognitive architecture: a critical analysis. *Cognition, 28*, 3-71.

Foulis, D. J. (1999). A half century of quantum logic—what have we learned? In D. Aerts, & J. Pykacz (Eds.), *Quantum Structures and the Nature of Reality : The Indigo Book of Einstein meets Magritte*. Dordrecht: Kluwer.

Foulis, D. J., Piron, C., & Randall, C. H. (1983). Realism, operationalism, and quantum mechanics. *Foundations of Physics, 13*, 813-841.

Foulis, D. J., & Randall, C. H. (1972). Operational statistics I: Basic concepts. *Journal of Mathematical Physics, 13*, 1667-1675.

Franco, R. (2007a). The conjunction fallacy and interference effects. *Arxiv preprint arXiv:0708.3948*.

Franco, R. (2007b). Quantum mechanics and rational ignorance. *Arxiv preprint physics/0702163*.

Franco, R. (2007c). Risk, ambiguity and quantum decision theory. *Arxiv preprint arXiv:0711.0886*,

Gabora, L., & Aerts, D. (2002). Contextalizing concepts using a mathematical generalization of the quantum formalism. *Journal of Experimental and Theoretical Artificial Intelligence*.

Gigerenzer, G. (1997). Bounded rationality models of fast and frugal inference. *Swiss Journal of Economics and Statistics, 133*, 201-218.

Gigerenzer, G., & Selten, R. (2001). *Bounded Rationality: The Adaptive Toolbox*. Cambridge, MA MIT Press.

Gleason, A. M. (1957). Measures on the closed subspaces of a Hilbert space. *Journal of Mathematics and Mechanics, 6*, 885-894.

Grice, P. (1989). *Studies in the way of words*. Cambridge Mass.: Harvard University Press.

Haag, R. (1992). *Local Quantum Physics: Fields, Particles, Algebras*. Berlin: Springer.

Hájek, A. (2012). Interpretations of Probability. *The Stanford Encyclopedia of Philosophy, available from http://plato.stanford.edu/archives/win2012/entries/probability-interpret/, Winter 2012 Edition*.

Halpern, J. Y. (2003). *Reasoning about Uncertainty*. Cambridge, MA: MIT Press.

Hampton, J. (1987). Inheritance of attributes in natural concept conjunctions. *Memory and Cognition, 15*, 55-71.

Hampton, J. (1988a). Disjunction of natural concepts. *Memory and Cognition, 16*(579-591).

Hampton, J. (1988b). Overextension of Conjunctive Concepts: Evidence for a Unitary Model of Concept Typicality and Class Inclusion. *Journal of Experimental Psychology: Learning, Memory, and Cognition, 14*(1), 12-32.

Hampton, J. (2007). Typicality, graded membership, and vagueness. *Cognitive Science, 31*, 355-384.

Hao, B.-L. (1989). *Elementary Symbolic Dynamics and Chaos in Dissipative Systems*. Singapore: World Scientific.

Heisenberg, W. (1944). *Die physikalischen Prinzipien der Quantentheorie*. Leipzig: S. Hirzel.

Hinzen, W. (2000). Minimalism. In R. Kempson, T. Fernando, & N. Asher (Eds.), *Handbook of Philosophy of Science 14: Philosophy and Linguistics* (pp. 93-142). Amsterdam: Elsevier.

Holland, S. S. (1995). Orthomodularity in infinite dimensions; a theorem of M. Solèr. *Bull. Amer. Math. Soc., 32*, 205-234. Also available as arXiv:math/9504224.

Howard, D. (2004). Who invented the "Copenhagen Interpretation"? A study in mythology. *Philosophy of Science, 71*(5), 669-682.

James, W. (1890). *The Principles of Psychology*. New York/London: Holt and Macmillan.

Jammer, M. (1989). *The Conceptual development of quantum mechanics* American Inst. Phys.

Jauch, J. M. (1968). *Foundations of Quantum Mechanics*. Reading, Massachusetts: Addison-Wesley.

Jenca, G. (2001). Blocks of homogeneous effect algebras. *Bulletin of the Australian Mathematical Society, 64*(1), 81-98.

Jiménez-Montaño, M. A., Ebeling, W., Pohl, T., & Rapp, P. E. (2002). Entropy and complexity of finite sequences as fluctuating quantities. *BioSystems, 64*, 23-32.

Johnson, E. J., Haubl, G., & Keinan, A. (2007). Aspects of endowment: A query theory of value construction. *Journal of Experimental Psychology: Learning, Memory and Cognition, 33*(3), 461-473.

Johnson, J. G., & Busemeyer, J. R. (2010). Decision making under risk and uncertainty. *Wiley Interdisciplinary Reviews: Cognitive Science, 1*(5), 736-749.

Kahneman, D., & Tversky, A. (1979). Prospect theory: An analysis of decision under risk. *Econometrica: Journal of the Econometric Society*, 263-291.

Kamp, H., & Partee, B. (1997). Prototype theory and compositionality. *Cognition, 64*(2), 189-206.

Khrennikov, A. Y. (2003a). Quantum-like formalism for cognitive measurements. *BioSystems, 70*(3), 211-233.

Khrennikov, A. Y. (2003b). Reconstruction of quantum theory on the basis of the formula of total probability. *Arxiv preprint quant-ph/0302194*,

Khrennikov, A. Y. (2006). Quantum-like brain: "Interference of minds". *BioSystems, 84*(3), 225-241.

Khrennikov, A. Y. (2010). *Ubiquitous quantum structure: From psychology to finance*. Berlin, New York: Springer.

Kitto, K. Why quantum theory? In P. Bruza, W. Lawless, C. van Rijsbergen, D. Sofge, B. Coecke, & S. Clark, Eds. Proceedings (Eds.), *Second Quantum Interaction Symposium, Oxford, 2008* (pp. 11-18)

Kuhn, T. S. (1996). *The structure of scientific revolutions*. Chicago: University of Chicago press.

La Mura, P. (2009). Projective expected utility. *Journal of Mathematical Psychology, 53*(5), 408-414.

Large, E. W. (2010). A dynamical systems approach to musical tonality. In R. Huys, & V. K. Jirsa (Eds.), *Nonlinear Dynamics in Human Behavior* (pp. 193-211). Berlin, New York: Springer.

Lerdahl, F. (2001). *Tonal Pitch Space*. NewYork: Oxford University Press.

Lewis, D. (1983). *Philosophical Papers Volume I*. Oxford: Oxford University Press.

Lind, D., & Marcus, B. (1995). *An Introduction to Symbolic Dynamics and Coding*. Cambridge, UK: Cambridge University Press.

Logan, G. D. (1988). Toward an instance theory of automatization. *Psychological Review, 95*, 492-527.

Lüders, G. (1951). Über die Zustandsänderung durch den Meßprozeß. *Annalen der Physik, 8*, 322-328.

Mackey, G. (1963). *The Mathematical Foundations of Quantum Mechanics*. New York: W.A. Benjamin.

Matsumoto, K. (1997). On C*-algebras associated with subshifts. *International Journal of Mathematics, 8*, 357-374.

Mazzola, G. (2002). *The topos of music: geometric logic of concepts, theory, and performance*. Basel: Birkhauser.

Montague, R. (1970). Universal grammar. *Theoria, 36*, 373-398.

Moore, C. (1990). Unpredictability and undecidability in dynamical systems. *Physical Review Letters, 64*, 2354-2357.

Moore, D. W. (2002). Measuring New Types of Question-Order Effects: Additive and Subtractive. *The Public Opinion Quarterly, 66*(1), 80-91.

Moore, R. C. (1988). Autoepistemic logic. In P. Smets (Ed.), *Non-Standard Logics for Automated Reasoning* (pp. 105-136). New York: Academic Press.

Müller, J. (1970). *Grundlagen der Systematischen Heuristik, Schriften zur Sozialistischen Wirtschaftsführung*. Berlin: Dietz-Verlag.

Murdoch, D. (1987). *Niels Bohr's philosophy of physics*: Cambridge University Press.

Niestegge, G. (2008). An Approach to Quantum Mechanics via Conditional Probabilities. *Foundations of Physics, 38*, 241-256.

Ott, E. (1993). *Chaos in Dynamical Systems*. New York: Cambridge University Press.

Piron, C. (1972). Survey of General Quantum Physics. *Foundations of Physics, 2*(4), 287-314.

Piron, C. (1976). *Foundations of quantum physics*. Reading, Mass.: WA Benjamin, Inc.

Pitowsky, I. (1989). *Quantum Probability - Quantum Logic*. Berlin, New York: Springer.

Popper, K. R. (1959). The Propensity Interpretation of Probability. *British Journal of the Philosophy of Science, 10*, 25-42.

Pothos, E. M., & Busemeyer, J. R. (2009). A quantum probability explanation for violations of 'rational' decision theory. *Proceedings of the Royal Society B: Biological Sciences, 276*.

Pothos, E. M., & Busemeyer, J. R. (2013). Quantum principles in psychology: The debate, the evidence, and the future. *Behavioral and Brain Sciences, 36*(3), 310-327.

Primas, H. (1990). Mathematical and philosophical questions in the theory of open and macroscopic quantum systems. In A. I. Miller (Ed.), *Sixty-two Years of Uncertainty: Historical, Philosophical and Physics Inquiries into the Foundation of Quantum Mechanics* (pp. 233-257): Plenum Press.

Primas, H. (2007). Non-Boolean Descriptions for Mind-Matter Problems. *Mind & Matter, 5*(1), 7-44.

Rabinovich, M. I., Huerta, R., & Laurent, G. (2008). Transient dynamics for neural processing. *Science, 321*(48 - 50).

Randall, C. H., & Foulis, D. J. (1973). Operational statistics II: Manuals of operations and their logics. *Journal of Mathematical Physics, 14*, 1472-1480.

Ratcliff, R. (1978). A theory of memory retrieval. *Psychological Review, 85*, 59 - 108.

Ratcliff, R., & Rouder, J. N. (1998). Modeling response times for two-choice decisions. *Psychological Science, 9*, 347 - 356.

Reed, A. V. (1973). Speed-accuracy trade-off in recognition memory. *Science, 181*, 574 - 576.

Richman, F., & Bridges, D. S. (1999). A constructive proof of Gleason's theorem. *Journal of Functional Analysis 162, 287 (1999), http://dx.doi.org/10.1006/jfan.1998.3372., 162*(287).

Rubinstein, A. (1998). *Modeling Bounded Rationality*. Cambridge, MA: MIT Press.

Russell, S., & Norvig, P. (1995). *Artificial Intelligence - A Modern Approach*. New Jersey: Prentice Hall.

Savage, L. J. (1954). *The Foundations of Statistics*. New York, NY: Wiley.

Schuman, H., & Presser, S. (1981). *Questions and answers in attitude surveys: Experiments in question form, wording, and context*. New York: Academic Press.

Searle, J. R. (1980). Minds, Brains, and Programs. *Behavioral and Brain Sciences, 3*, 417-457.

Searle, J. R. (1998). *Mind, language and society: Philosophy in the real world*. Cambridge, Mass.: Cambridge Univ Press.

Simon, H. A. (1955). A behavioral model of rational choice. *The Quarterly Journal of Economics, 69*(1), 99-118.

Smets, S. (2011). Logic and Quantum Physics. *Journal of the Indian Council of Philosophical Research*.

Solér, M. P. (1995). Characterization of Hilbert spaces by orthomodular spaces. *Communications in Algebra, 23*(1), 219-243.

Spenader, J., & Blutner, R. (2007). Compositionality and Systematicity. In G. Bouma, I. Krämer, & J. Zwarts (Eds.), *Cognitive Foundations of Interpretation*. Amsterdam: Verhandelingen, Afd. Letterkunde, Nieuwe Reeks.

Stapp, H. P. (1972). The Copenhagen interpretation. *AJP*, 1098-1116.

Sternefeld, W. (2012). Wo stehen wir in der Grammatiktheorie? Bemerkungen anlässlich eines Buchs von Stefan Müller. Available from http://www.s395910558.online.de/Downloads/Rezension-15-lang.pdf. *Seminar für Sprachwissenschaft*. Tübingen.

Stokhof, M., & van Lambalgen, M. (2011). Abstractions and idealisations: The construction of modern linguistics. *Theoretical Linguistics, 37*(1-2), 1-26.

Streater, R. F., & Wightman, A. S. (1964). *PCT, Spin & Statistics, and all that*. New York: Benjamin.

Sudman, S., & Bradburn, N. M. (1982). *Asking questions*. San Francisco: Jossey-Bass Inc Pub.

Suppes, P., Krantz, D. H., Luce, R. D., & Tversky, A. (1989). *Foundations of measurement: Geometrical, threshold, and probabilistic representations*. London: Academic Press.

Tversky, A. (1977). Features of similarity. *Psychological Review, 84*, 327-352.

Tversky, A., & Fox, C. R. (1995). Weighing risk and uncertainty. *Psychological Review, 102*(2), 269-283.

Tversky, A., & Kahneman, D. (1983). Extension versus intuitive reasoning: The conjunction fallacy in probability judgment. *Psychological Review, 90*(4), 293-315.

Tversky, A., & Kahneman, D. (1992). Advances in prospect theory: Cumulative representation of uncertainty. *Journal of Risk and uncertainty, 5*(4), 297-323.

Tversky, A., & Shafir, E. (1992). The disjunction effect in choice under uncertainty. *Psychological Science, 3*(5), 305-309.

Tymoczko, D. (2011). *A Geometry of Music*: Oxford University Press.

Vedral, V. (2006). *Introduction to Quantum Information Science*. New York: Oxford University Press.

von Neumann, J. (1932). *Mathematische Grundlagen der Quantenmechanik. Transl. by R. Beyer (Princeton: Princeton University Press, 1955).* Heidelberg: Springer.

Von Neumann, J., & Morgenstern, O. (1944). *The theory of games and economic behavior*. Princeton, NJ: Princeton University Press.

Wang, Z., & Busemeyer, J. R. (2013). A quantum question order model supported by empirical tests of an a priori and precise prediction. *Topics in Cognitive Science, 5*(4), 689-710.

Weber, E. U., & Johnson, E. J. (2006). Constructing preferences from memories. In S. Lichtenstein, & P. Slovic (Eds.), *The Construction of preference* (pp. 397-410). New York, NY: Cambridge University Press.

Weber, E. U., & Johnson, E. J. (2009). Mindful judgement and decision making. *Annual Review of Psychology, 60*, 53-85.

Wickelgren, W. A. (1977). Speed-accuracy tradeoff and information processing dynamics. *Acta Psychologica, 41*, 67 - 85.

Wilce, A. (2009). Test spaces. In K. Engesser, D. M. Gabbay, & D. Lehmann (Eds.), *Handbook of Quantum Logic and Quantum Structures* (pp. 443-549). Amsterdam: Elsevier.

Wilce, A. (2012). Quantum Logic and Probability Theory. In E. N. Zalta (Ed.), *The Stanford Encyclopedia of Philosophy*. Stanford: CSLI, available from http://plato.stanford.edu/archives/fall2012/entries/qt-quantlog/.

Yukalov, V. I., & Sornette, D. (2010). Decision theory with prospect interference and entanglement. *Theory and Decision, 70*, 283-328.

Yukalov, V. I., & Sornette, D. (2013). Quantum probabilities of composite events in quantum measurements with multimode states. *Laser Physics, 23*(10), doi:doi:10.1088/1054-660X/23/10/105502.